%% file: paper_conext_2023.tex
\documentclass[sigconf,10pt]{acmart}
\makeatletter
\def\@ACM@checkaffil{
    \if@ACM@instpresent\else
    \ClassWarningNoLine{\@classname}{No institution present for an affiliation}%
    \fi
    \if@ACM@citypresent\else
    \ClassWarningNoLine{\@classname}{No city present for an affiliation}%
    \fi
    \if@ACM@countrypresent\else
        \ClassWarningNoLine{\@classname}{No country present for an affiliation}%
    \fi
}
\makeatother
\renewcommand\footnotetextcopyrightpermission[1]{} 
\acmConference{}{}{}

\AtBeginDocument{%
  \providecommand\BibTeX{{%
    \normalfont B\kern-0.5em{\scshape i\kern-0.25em b}\kern-0.8em\TeX}}}

\usepackage[english]{babel}
\usepackage{blindtext}
\usepackage{tikz}
\usepackage{amsmath}
\usepackage{enumitem}
\usepackage{booktabs}
\usepackage[font=small,labelfont=bf,textfont=it]{caption}
\usepackage[hang,scriptsize,tight,nooneline]{subfigure}
\usepackage{gensymb}
\usepackage{outlines}
\usepackage{dirtytalk}
\usepackage{multirow}
\usepackage{graphicx}
\usepackage{subcaption}
\usepackage{sidecap}
\newcommand{\cut}[1]{}
\newcommand{\projectname}{\texttt{LEOScope}}
\newcommand{\starlinksubreddit}{\texttt{r/Starlink}}

\newcommand{\parab}[1]{\vspace{0.03in}\noindent{\bf #1}}

\newcommand{\db}{\textcolor{black}}

\newcommand{\ns}{\textcolor{orange}}

\definecolor{lightgray}{gray}{0.92}
\newcommand\greybox[1]{%
\vspace{6pt}%
      \par{\centering\colorbox{lightgray}{%
              \begin{minipage}{3.3in}#1\end{minipage}%
                        }%
                                    \vskip 2pt%
                                    \vspace{1pt}%
                                                }}

\begin{document}

\title{\huge T$3$P: Demystifying Low-Earth Orbit Satellite Broadband}

\author{%
  \fontsize{12pt}{12pt}\selectfont{}Shubham Tiwari$^{1}$,
  \fontsize{12pt}{12pt}\selectfont{}Saksham Bhushan$^{1}$,
  \fontsize{12pt}{12pt}\selectfont{}Aryan Taneja$^1$,
  \fontsize{12pt}{12pt}\selectfont{}Mohamed Kassem$^2$,
  \fontsize{12pt}{12pt}\selectfont{}Cheng Luo$^3$,\\
  \fontsize{12pt}{12pt}\selectfont{}Cong Zhou$^4$,
  \fontsize{12pt}{12pt}\selectfont{}Zhiyuan He$^3$,
  \fontsize{12pt}{12pt}\selectfont{}Aravindh Raman$^5$,
  \fontsize{12pt}{12pt}\selectfont{}Nishanth Sastry$^2$,
  \fontsize{12pt}{12pt}\selectfont{}Lili Qiu$^3$,
  \fontsize{12pt}{12pt}\selectfont{}Debopam Bhattacherjee$^1$
}
\affiliation{%
\vspace{0.1in}
  \fontsize{10pt}{12pt}\selectfont{}%
  $^1$Microsoft Research India, $^2$University of Surrey, UK, $^3$Microsoft Research Asia,\\ $^4$Shanghai Jiao Tong University, $^5$Telefonica Research, Spain
\vspace{0.1in}
}

\renewcommand{\shortauthors}{Trovato and Tobin, et al.}

\begin{abstract}
\textit{The Internet is going through a massive infrastructural revolution with the advent of low-flying satellite networks, $5/6$G, WiFi$7$, and hollow-core fiber deployments. While these networks could unleash enhanced connectivity and new capabilities, it is critical to understand the performance characteristics to efficiently drive applications over them. Low-Earth orbit (LEO) satellite mega-constellations like SpaceX Starlink aim to offer broad coverage and low latencies at the expense of high orbital dynamics leading to continuous latency changes and frequent satellite hand-offs.}

\textit{This paper aims to quantify Starlink's latency and its variations and components using a real testbed spanning multiple latitudes from the North to the South of Europe. We identify tail latencies as a problem. We develop predictors for latency and throughput and show their utility in improving application performance by up to $25\%$. We also explore how transport protocols can be optimized for LEO networks and show that this can improve throughput by up to $115\%$ (with only a $5\%$ increase in latency). Also, our measurement testbed with a footprint across multiple locations offers unique trigger-based scheduling capabilities that are necessary to quantify the impact of LEO dynamics.}
\end{abstract}

\maketitle
\pagestyle{plain}

\input{intro.tex}

\input{motivation.tex}

    
\input{latency_measurements.tex}
\input{video-streaming-predictor.tex}
\input{bbr2_parameter_tune.tex}

\input{testbed-2.tex}

\input{discussions.tex}

\section{Conclusion}
In this paper, we deep dive into the performance of Starlink tail latencies via a distributed testbed spanning multiple latitudes from Edinburgh in North Europe to Barcelona in the South. We find that although Starlink provides high throughput at low latencies, the tail of the latency distribution can be a problem, with the worst-case latency being $11$-$16\times$ the median. To help applications manage such huge network performance changes, we build predictors and show how applications such as video streaming can consume such predictions to obtain better performance (up to $25\%$). We then show that transport protocols such as BBRv$2$ can be fine-tuned to work better on LEO networks, obtaining an impressive $115\%$ improvement over the default parameters. These points to directions for future work on making networked applications better on LEO networks. 

\parab{Ethics. } 
We run active experiments on \projectname{}, with due permissions from the measurement client hosts, to collect all data. All measurement servers are either deployed by us on the public cloud or host publicly known Internet services (DNS and CDN). Hence, none of the IP addresses logged in our experiments raise PII concerns. For the motivation plots in \S\ref{sec:motivation}, we only use publicly available data and screenshots gathered using standard APIs. Hence this work does not raise any ethical concerns.

\bibliographystyle{ACM-Reference-Format}
\bibliography{reference}

\end{document}

%% file: intro.tex
\section{Introduction}
\label{sec:intro}

There has been a recent revolution in network connectivity with multiple big players like SpaceX (Starlink)~\cite{starlink}, OneWeb~\cite{oneweb}, Telesat (Lightspeed)~\cite{telesat}, Amazon (Kuiper)~\cite{kuiper}, and others starting to offer Internet services across the globe leveraging significant reductions~\cite{launchcost} in satellite launch costs. Most of these constellations target Low-Earth Orbit or LEO~\cite{leo_wiki} heights ($<${}$2$,$000$~km) to reduce latencies by order(s) of magnitude compared to geosynchronous or GEO~\cite{geo_wiki} satellites which require an altitude of $35$,$786$~km. 

While there is much excitement that LEO could leverage lower latencies to deliver traffic for fast interactive networked applications, this configuration also necessitates satellites flying at blazing speeds offering connectivity windows of only a few minutes to ground terminals. Such transient windows lead to frequent hand-offs at the ground terminal between satellites, which can potentially have a negative effect on applications. 

In this paper, we shed light on 
the network dynamics of the \textit{real} SpaceX Starlink LEO network and the ramifications of significant latency fluctuations on transport and application layers. We found that although Starlink provides high throughput at low latencies, the tail of the latency distribution can be a problem, with the worst case latencies being 11-16x the median latencies, which can greatly affect application performance.

We explore two parallel threads to address some of the performance bottlenecks we find. First, we construct predictive models tailored to LEO networks, capable of accurately estimating latency and throughput. We treat the prediction task as a time series estimation and regression problem and develop models using XGBoost and LSTM and features like the Starlink terminal orientation (extracted from the telemetry data exposed by Starlink terminals via a gRPC API), current satellite positions (from Celestrak~\cite{celestrak}), and past telemetry data from our continuous experiments. The best-performing latency (throughput, resp.) prediction model achieves a MAPE (Mean Average Percentage Error) of $3.65\%$ ($19.2\%$).  We augment a cutting-edge video streaming Adaptive Bitrate (ABR) algorithm, RobustMPC~\cite{yin2015control}, to consume throughput predictions from our custom-built model. This integration of LEO awareness improves video streaming Quality of Experience (QoE; Pensieve)~\cite{pensieve_qoe} by up to $25\%$ demonstrating how cross-layer insights could help applications improve performance over these dynamic networks.

In the second thread of exploration, we quantify and augment end-to-end transport performance over Starlink LEO. Based on our understanding of the impact of the network dynamics on transport performance, we pursue an intuitive $2$-dimensional parameter space exploration of BBRv$2$~\cite{cardwell2019bbrv2} based on the expectation that (a) LEO latencies fluctuate due to the continous motion of satellites  as well as abruptly due to hand-offs between satellites, thus necessitating the need to probe for the minimum latency more often, and (b) congestion control should be robust to non-congestive and higher loss~\cite{kassem2022browser,starlink_loss} over these networks. Our parameter search identifies BBRv$2$ configurations that offer $115\%$ higher throughput than the default setting with minimal inflation in latency. 

For running all our measurement and performance optimization experiments, we built a testbed, called \projectname{}, consisting of measurement clients behind Starlink terminals, measurement servers on Azure, and an orchestrating service deployed on Azure for scheduling experiments across nodes. \projectname{} consists of $5$ measurement clients (behind `pizza-box sized' Starlink terminals) spread across $2$ countries and measurement servers on Azure deployed geographically close to the clients. Also, \projectname{} offers two useful features (discussed later), trigger-based scheduling and scavenger mode, focused on the LEO dynamics and the scalability of the testbed. \db{As part of this work, we have released the \projectname{} code~\cite{leoscope_code}.}

\parab{}Our key contributions are as follows\footnote{The title of the paper, T$3$P, is influenced by these key contributions. Note that T$3$P is a solution stack discussed in \S\ref{subsec:t3p_stack}.}:
\begin{itemize}

    \item \textbf{\underline{T}elemetry:} We conduct SpaceX Starlink performance measurement experiments across different vantage points. We focus on measuring the latency to services on the Web that are critical -- DNS and CDN servers. Our latency dissection analysis shows how the bent pipe (client to Starlink terminal to ground station to the Internet and vice versa) contributes to the path latency.

    \item \textbf{\underline{P}redictors:} Our custom-built LEO performance predictors demonstrate useful QoE enhancement opportunities when plugged into applications.

    \item \textbf{\underline{T}ransport:} Our BBRv$2$ $2$-dimensional parameter tuning experiment demonstrates performance improvement opportunities by making end-to-end congestion control aware of the LEO dynamics. This opens up a research space to explore clean-slate congestion control design over Starlink-like LEO networks.

    \item \textbf{\underline{T}riggers:} We built a distributed testbed, \projectname{}, to run our experiments which offers trigger-based scheduling and scavenger mode useful for catching LEO performance characteristics and scaling the testbed respectively. Triggers could be used to probe for arbitrary events of interest -- LEO latency inflation, bad weather, terminal movements, satellite positions, etc.

    
\end{itemize}

%% file: motivation.tex
\section{Background \& Motivation}
\label{sec:motivation}

In this section, we discuss the current state, goals, and dynamics of LEO broadband networks. SpaceX Starlink satellites deployed at $\sim${}$550$~km orbital heights fly at $7.5$~km/s~\cite{bhattacherjee2019network}, more than $20\times$ faster than sound. While we have seen limited mobility of infrastructure and/or clients, thousands of routers/switches flying in the sky at these incredible speeds make LEO networks interesting to study.

\subsection{LEO Broadband Deployments} 

 While previous efforts such as the Iridium constellation~\cite{iridium} also operated in the Low Earth Orbit, LEO broadband did not take off due to the parallel cellular communication boom~\cite{iridium_vs_cellular} and the exorbitant cost~\cite{launchcost} of sending satellite payloads to space at that time. Since then the space industry has seen multiple major advances in technology -- reusable boosters~\cite{reusable_booster} reducing launch costs by order(s) of magnitude, compact and cheap satellite design~\cite{cubesat}, inter-satellite link tracking and laser communication~\cite{intersat_laser}, etc. Such developments have now made it economically viable to launch large Low-Earth orbit mega-constellations, consisting of thousands to tens of thousands of satellites, as proposed/deployed by multiple players like SpaceX, OneWeb, Amazon, Telesat, and others. These dense constellations, deployed at much lower heights than GEO satellites, promise coverage across large areas of the world, at significantly lower latencies than GEO. This approach requires a high density of satellites to ensure that each user terminal across the globe can connect to one at any point in time.

\begin{figure*}[t]
  \centering
  \subfigure[]{\label{fig:trends}
    \includegraphics[width=0.350\textwidth]{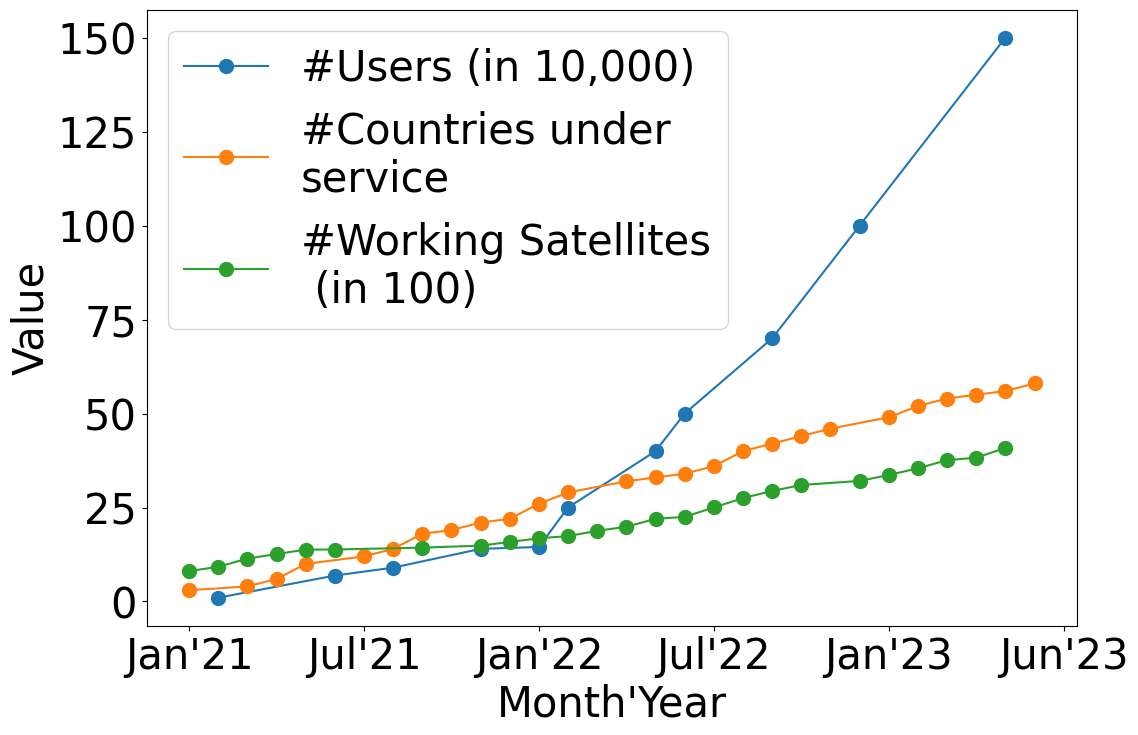}}
  \subfigure[]{\label{fig:redditspeed}
    \includegraphics[width=0.285\textwidth]{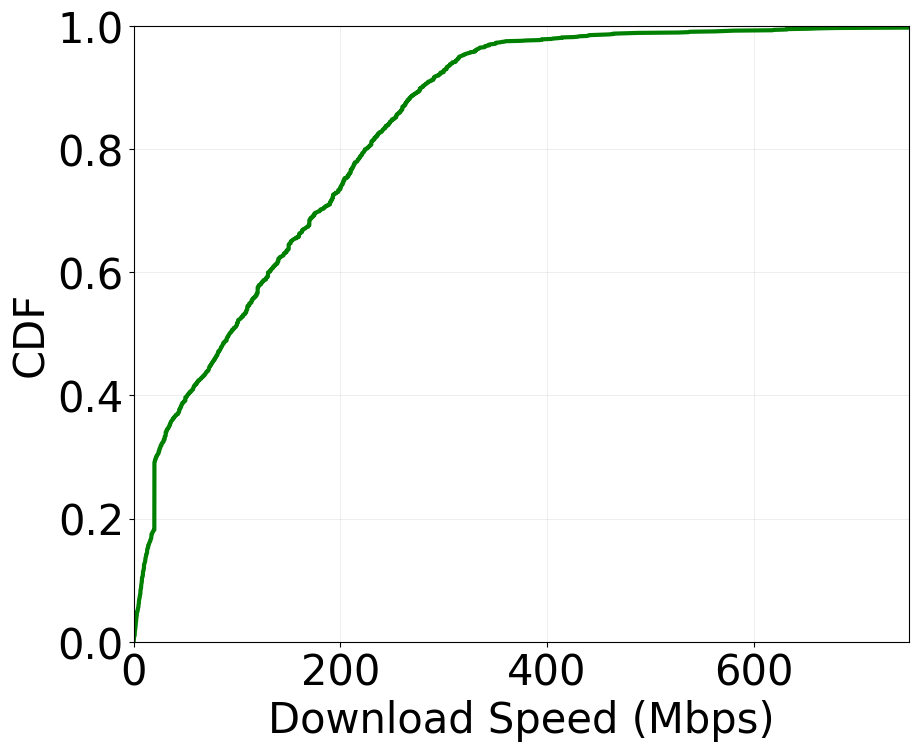}}
    \subfigure[]{\label{fig:redditlatency}
    \includegraphics[width=0.295\textwidth]{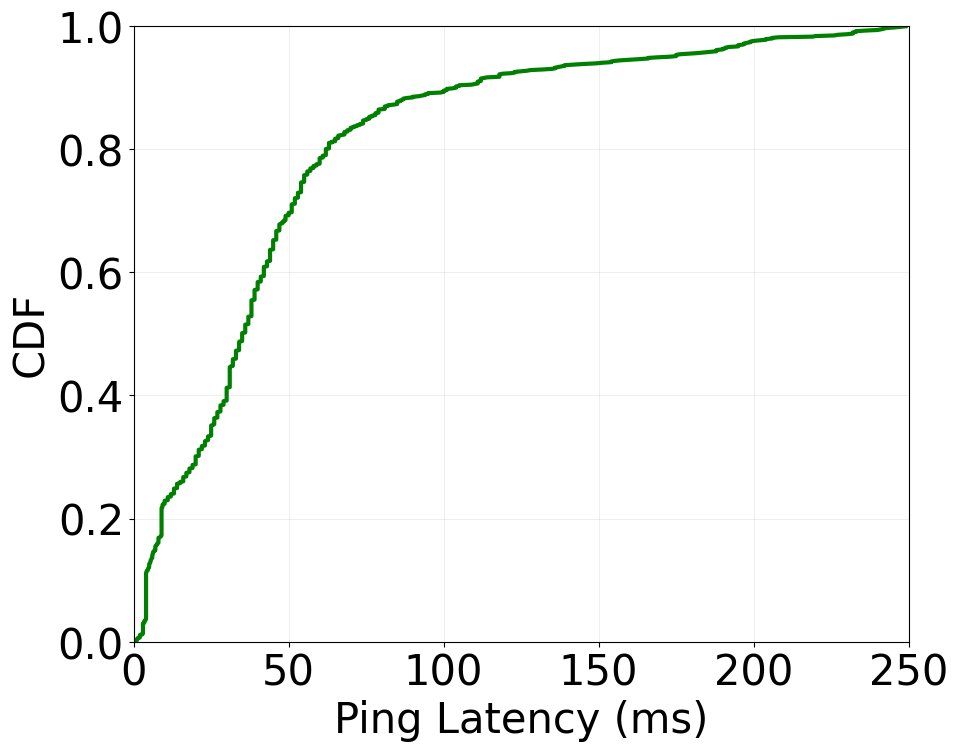}}
  \caption{(a) The number of Starlink satellites and the number of serviceable countries are increasing $\sim$linearly over time. Interestingly, the number of users is growing super-linearly. Data collected from publicly available sources~\cite{launch_data_1, launch_data_2, launch_data_3, starlink_wiki, beta_10K, beta_69K, beta_90K, beta_140K, beta_250K, beta_400K, beta_500K, beta_700K, beta_1M, beta_1.5M}. (b) and (c) shows the download speeds and ping latencies resp. reported by Starlink users on Reddit.}
  \label{fig:redditmotivation}
\end{figure*}

SpaceX Starlink is leading the front and has already deployed $\sim${}$4$,$000+$\cite{starlink_wiki} LEO satellites and has started offering Internet connectivity services in $58$~\cite{starlink_wiki} countries across the globe. 
Fig.~\ref{fig:trends} shows how these numbers are increasing over time since Jan'$21$. Interestingly, user adoption is growing super-linearly with the number of Starlink users touching $1.5$ million~\cite{starlink_wiki} recently. OneWeb has also deployed $500+$~\cite{celestrak} LEO satellites already.


\subsection{LEO Broadband Goals} 

\parab{Coverage goal} SpaceX’s stated goal~\cite{elon_claim} is ``to have the \textit{majority of long distance Internet traffic} go over this (Starlink) network''. While this is yet to be achieved, as Fig.~\ref{fig:trends} shows, the number of Starlink users has increased at an incredibly fast pace, and Starlink offerings have expanded across many countries in just a few years. Starlink has also engaged to offer continued connectivity in conflict and disaster-stricken regions. It is straightforward to install user terminals anywhere on earth and connect to satellites overhead -- LEO orbital geometry makes sure that providers need to deploy enough satellites uniformly between servicing north and south latitudes to offer continuous connectivity. Interestingly, this is a highly democratic and inclusive design of a globally spanning network, enforced by the Earth's gravity and rotation.

Similar to SpaceX, OneWeb states~\cite{oneweb_claim} that it aims to ``enable high-speed, low latency connectivity for governments, businesses, and communities \textit{everywhere} around the world''. Amazon's Kuiper aims~\cite{kuiper_claim} to ``provide low-latency, high-speed broadband connectivity to unserved and underserved communities \textit{around the world}''. 



\parab{Latency goal} As can be seen in the above publicly stated goals, most of the LEO broadband providers aim to offer `low-latency' Internet connectivity. SpaceX also claims the same~\cite{elon_latency_claim} for Starlink and argues that the latencies should be `good enough'~\cite{starlink_gaming_latency} for fast online gaming. Starlink currently offers bent-pipe connectivity with single satellites acting as intermediate hops between user terminals and terrestrial Internet gateways. SpaceX claims~\cite{starlink_gaming_latency} to be working on reducing the latencies further. With their latest generation satellites having multiple inter-satellite laser terminals being deployed currently, long-distance Internet latencies are set to improve~\cite{bhattacherjee2018gearing, handley2018delay} drastically compared to fiber.

\subsection{LEO Dynamicity} 
While LEO broadband aims to offer global low-latency connectivity, this comes at a performance cost. At LEO heights, satellites move at breakneck speeds ($7.5$~km/s at $550$~km height~\cite{bhattacherjee2019network}) thus offering a limited few-minute connectivity window to the user terminals underneath. While thousands of satellites in a constellation make sure that most or all of the user terminals get at least a few satellite options to connect to at any point in time, such extreme dynamics lead to frequent hand-offs and latency changes. While the core infrastructure itself is mobile unlike in terrestrial cellular networks, LEO satellite trajectories are predefined. Based on this expectation, we aim to predict latency/throughput changes and hand-offs, with the goal of improving transport, traffic engineering, and application performance over LEO networks.

\begin{figure*}[tbh]
  \centering
  \subfigure[]{\label{fig:dns_cloud_lat_cdf}
    \includegraphics[width=0.32\textwidth]{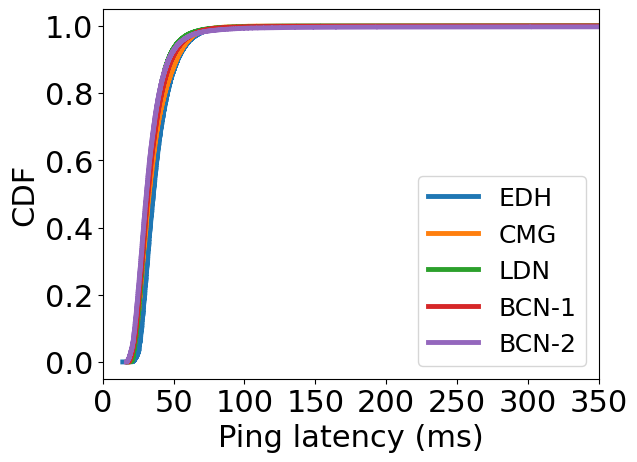}}
  \subfigure[]{\label{fig:bcn-1_cdn_cdf}
    \includegraphics[width=0.32\textwidth]{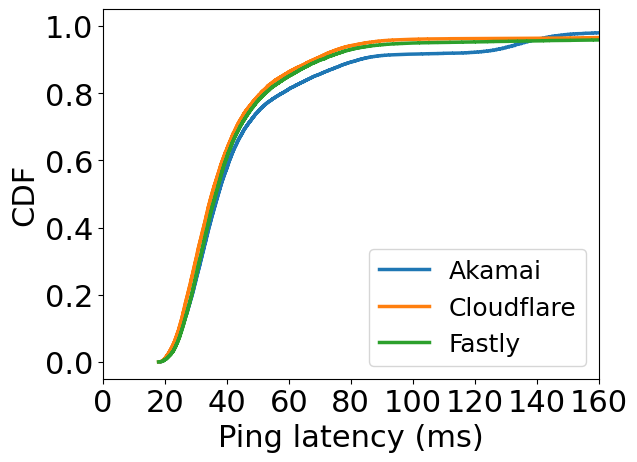}}
  \subfigure[]{\label{fig:akamai_cdf}
    \includegraphics[width=0.32\textwidth]{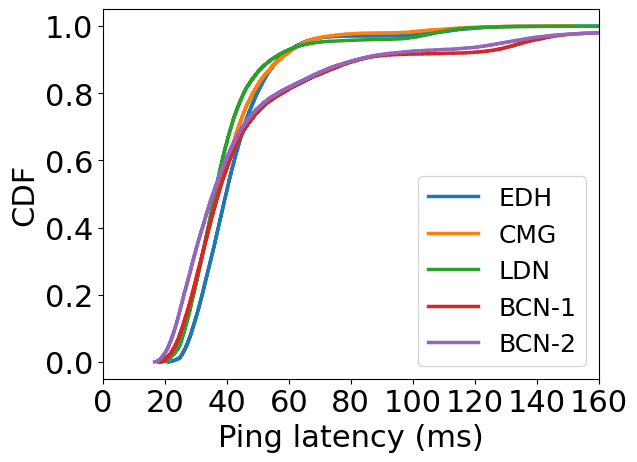}}
  \caption{Ping latency to (a) Google's $8$.$8$.$8$.$8$ from different locations, (b) Web domains fully hosted by popular CDNs from Barcelona, and (c) Web domains hosted by Akamai from different locations.}
  \label{fig:latency}
\end{figure*}

\subsection{LEO Performance Variations}
LEO broadband network performance could vary due to multiple key reasons -- LEO orbital dynamics as discussed above, the geometry of the constellations, weather having its toll on satellite-ground connectivity, continuous deployments, local demand changes, GEO arc-avoidance~\cite{starlink_FCC_spec_mod}, etc. To have a quick understanding of such variations, we mined social interactions on \starlinksubreddit{}~\cite{starlink_subreddit} subReddit. It is a thriving community with $370+$ posts and $5.5K+$ comments on average per week between Jan'$21$ and Dec'$22$\footnote{Newer data currently unavailable as Reddit is upgrading their APIs~\cite{reddit_upgrade}.} with redditors exhaustively discussing SpaceX Starlink bandwidth, latency, and outage, and sharing speed-test screenshots. We used Python Reddit API Wrapper (PRAW)~\cite{praw} to gather publicly available data from \starlinksubreddit{} for the above period, got rid of PII (Personally identifiable information), and used standard language and computer vision tools like Azure’s Cognitive Services (ACS)~\cite{azure_acs} and Natural Language Toolkit (NLTK)~\cite{nltk} to extract relevant information on network bandwidth, latency, loss, and outage from the posts, comments, and screenshots (PII removed). 
Fig.~\ref{fig:redditspeed} and \ref{fig:redditlatency} plot the CDFs of reported downlink speeds and ping latencies extracted from $\sim${}$1.7K$ public speed-test screenshots shared by Redditors during the period. The $95^{th}$ percentile speed (latency) is $\sim${}$3.4\times$($\sim${}$4.5\times$) the median value showing the significant variation in performance even though the constellation deployment is intended to be roughly uniform (going by the FCC filings~\cite{starlink_FCC_spec_mod, telesat_fcc_may20}) in the latitude bands being served. 
These observations call for a deeper understanding of LEO network performance at scale across space and time by running active experiments. 

%% file: latency_measurements.tex


\section{\underline{T}elemetry: Understanding LEO latency and its tail}
\label{sec:latency}

We start the T$3$P journey by taking a bird's-eye view of the latency offerings in SpaceX Starlink's LEO network. We quantify the latency to reach critical Web infrastructure that users expect low latency from --- public DNS and CDN servers. We dig deeper to dissect the components that constitute end-to-end latency and underscore the presence of persistent latency spikes, which could likely be attributed to satellite handovers.

\subsection{Experimental Setup}
While we have a detailed discussion on the \projectname{} testbed internals in \S\ref{sec:testbeds}, here we briefly touch upon the experiment setup to not stray too far from the focus of this work --- T$3$P. As \textit{\underline{T}riggers} are an inherent offering of the testbed, we discuss them later with \projectname{}. Our experiments are run on measurement clients behind Starlink terminals deployed in $2$ countries - Spain ($2${}$\times$; BCN-$1$ and BCN-$2$ in Barcelona) and UK ($3${}$\times$; EDH in Edinburgh, CMG in Cambridge, and LDN in London). These locations help us get an interesting spatial diversity. Starlink's first shell of satellites is known to be deployed at $53\degree$~\cite{starlink_FCC_spec_mod} inclination -- hence the density of satellites is highest around $53\degree$~N and S latitudes. Our measurement clients are distributed in a way that they cover the lower latitudes (BCN-$1$ \& BCN-$2$: $41.4\degree$~N), the area of high density (LDN: $51.5\degree$~N, CMG: $52.2\degree$~N), and latitudes further north of this dense ring (EDH: $55.9\degree$~N). The measurement servers are publicly known servers like DNS and CDN, or servers hosted by us on the public cloud geographically close to the measurement client deployments. All clients (servers, resp.) deployed by us are Standard Linux $5$.$15$.$0$ machines (VMs) with $2$ ($8$) v/CPUs, $8$ ($32$) GB of memory, and high-bandwidth network interfaces not rate limiting Starlink capacity. All measurements were run in June'$23$. The experiments were scheduled with \projectname{}'s orchestrator on Azure. Note that the same setup is used for running all experiments in this work.

\subsection{Latency to popular services}
\label{subsec:latency_transport_results}

First, we quantify Starlink's latency offering by reaching out to popular public Internet services like DNS and CDN from the measurement clients and gathering latency telemetry data. For this, we run ping and hping (TCP SYN-SYN/ACK handshake) latency probes from the measurement clients to public DNS and CDN servers, which are known to be well-distributed, at a cadence of $1$~second for $2$~hours, $3$ times a day for $7$ days. As shown in past work~\cite{bozkurt2017internet}, latency to the DNS and CDN services significantly contributes to the application/user-perceived latency and usually has a multiplicative effect on user-perceived delays, due to protocol overheads.


Fig.~\ref{fig:dns_cloud_lat_cdf} plots the CDFs of ping latencies to Google's anycast DNS servers ($8$.$8$.$8$.$8$)~\cite{google_dns}. The minimum and maximum median latencies to Google DNS are $30.8$~ms (BCN-2) and $35.7$~ms (EDH) respectively. Interestingly, at all of these locations, the latencies have long tails with the $99^{th}$ percentiles  being $>${}$2.4\times$ the median (maximum latency is $11$-$16\times$ the median). We observe similar trends with \texttt{hping} and with Cloudflare's DNS ($1$.$1$.$1$.$1$)~\cite{cloudflare_dns}. In the latency time-series analysis below, we deep-dive into the long-tail latencies observed over Starlink's LEO network.

For CDN services, we get the top Web domains from the Tranco list~\footnote{Available at https://tranco-list.eu/list/LY344 (generated on $04$ Dec, $2022$)}~\cite{tranco}, get the final redirected domains, filter domains (TLDs: com, net, edu, org, tv, io, site, gov), get corresponding IP addresses, and get the OrgName and NetName from \texttt{whois}~\cite{whois} registry to identify top $100$ sites fully served by (IP addresses belong to the CDN) $3$ popular CDN services \cite{cloudproviders} --- Akamai, Cloudflare, and Fastly. Note that these CDN services are the ones that fill up these $100$ full-site delivery slots fastest for the ordered Tranco list. Fig.~\ref{fig:bcn-1_cdn_cdf} plots the CDFs to these $3$ CDNs from BCN-$1$. All three CDN services have comparable performance for BCN-$1$ with the median latency being $37.3$~ms (Akamai), $35.2$~ms (Cloudflare), and $36.5$~ms (Fastly). However, latency to Akamai-hosted servers is $60\%$ higher at the $95^{th}$ percentile. 
Fig.~\ref{fig:akamai_cdf} shows the latencies to Akamai domains across all locations. Interestingly, we observed a clear separation in the latency performance across different measurement clients -- $95^{th}$ percentiles for EDH, CMG and LDN nodes in the UK are $64.6$\ns~ms, $65.1$~ms, $68.3$~ms respectively while the same for  BCN-$1$ and BCN-$2$ in Spain are $136$~ms and $129$~ms respectively. This translates to more than $2\times$ latency inflation for the latter nodes at the $95^{th}$ percentile. This could be a reflection of a mix of multiple factors including physical locations of gateway and CDN deployments.

The clear trend that we observe across all $3$ plots in Fig.~\ref{fig:latency} is that, while median latencies across all locations and services are low ($30$-$50$~ms), tail latencies are significantly higher, sometimes by an order of magnitude. We investigate the potential causes next.


\subsection{Investigating latency fluctuations}

\begin{figure*}[t]
  \centering
  \subfigure[]{\label{fig:latency_time_series}
    \includegraphics[width=0.32\textwidth]{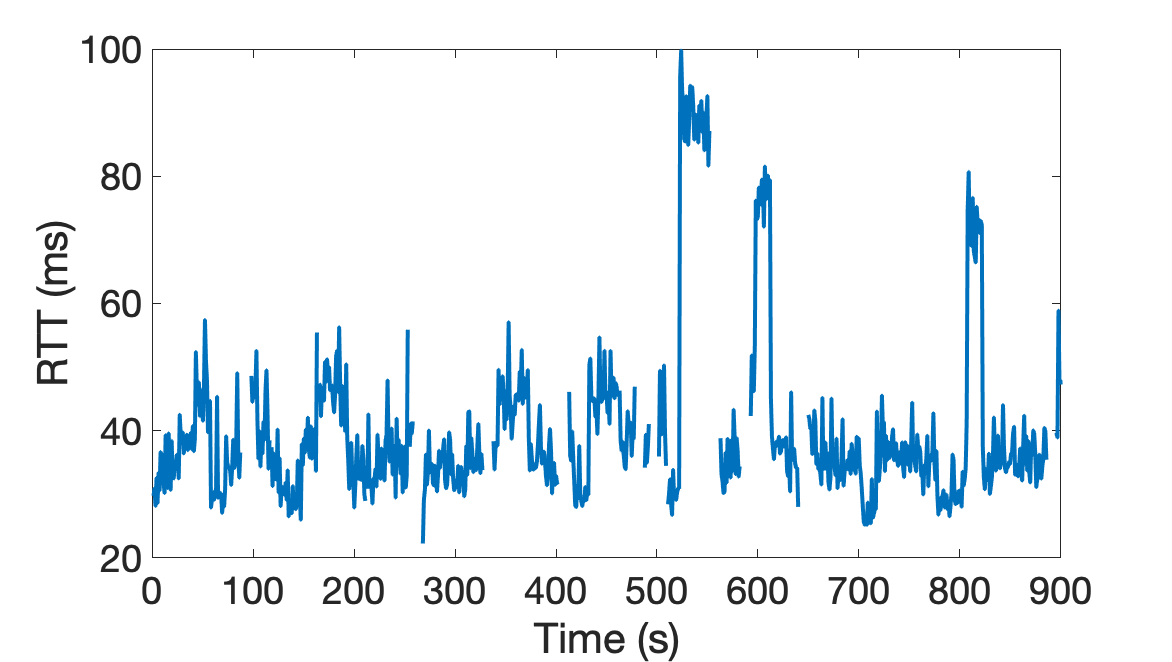}}
      \subfigure[]{\label{fig:leo_path}
    \includegraphics[width=0.32\textwidth]{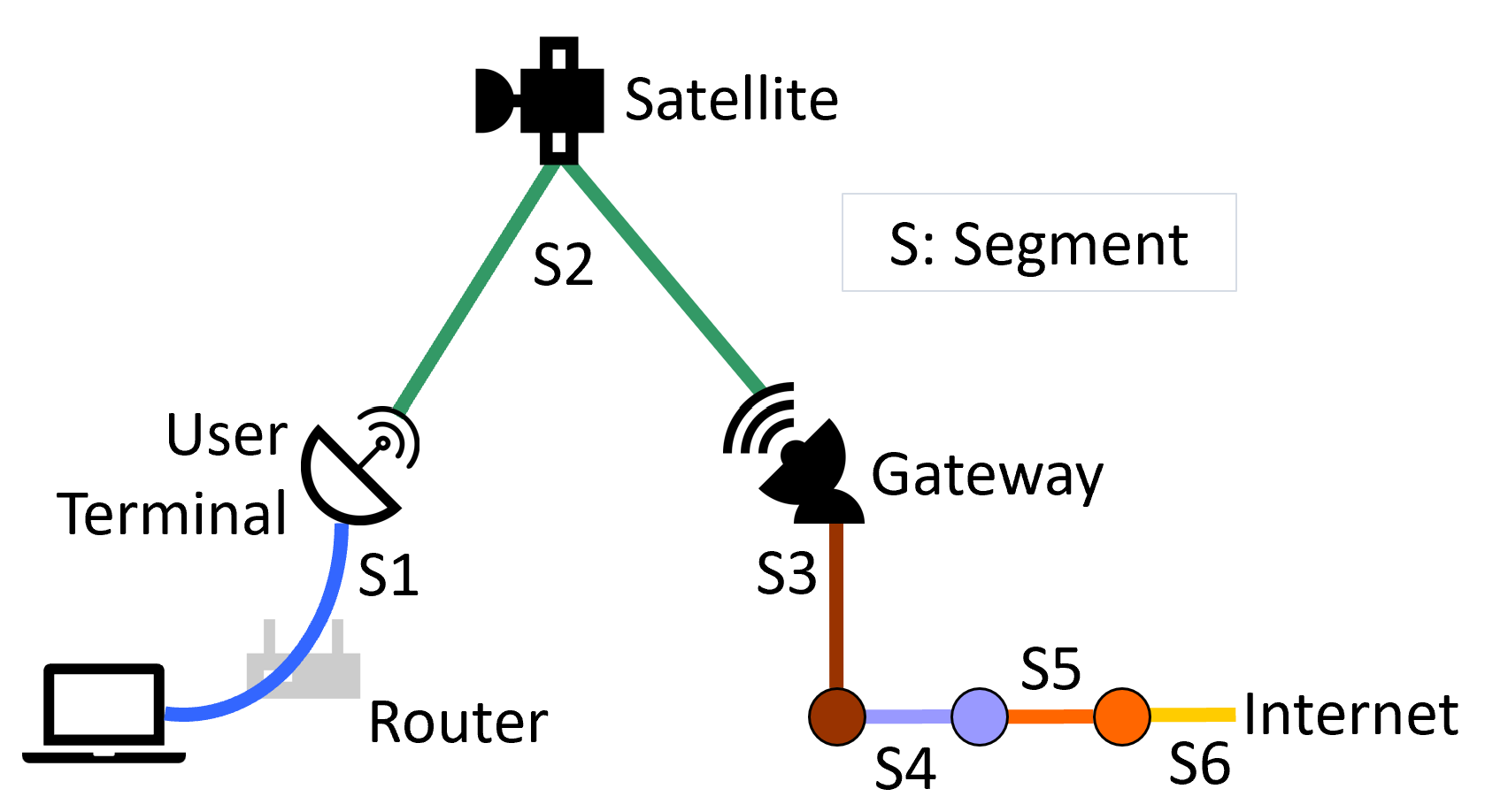}}
  \subfigure[]{\label{fig:latency_bar_plot}
    \includegraphics[width=0.32\textwidth]{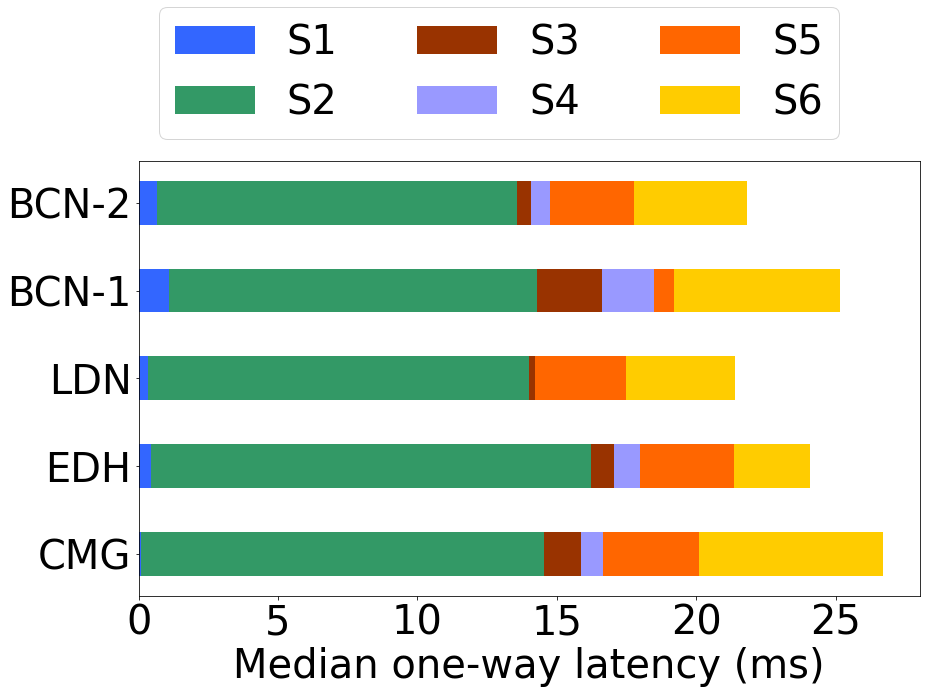}}
  \caption{(a) Latency varies over time with prominent spikes that persist for a while. (b) The individual color-coded segments of the path. S$2$ or the bent-pipe portion contributes the maximum latency. (c) The segment(S)-wise median latency for the end-to-end path to Google's DNS server.}
\end{figure*} 

In order to understand the long-tail latencies better, in Fig.~\ref{fig:latency_time_series}, we plot the latency time-series  for the ping latency to Google's $8$.$8$.$8$.$8$ from the BCN-$1$ node for a period of $15$~min picked at random from the measurement windows in Fig.~\ref{fig:dns_cloud_lat_cdf}. While the observed median latency is $35.2$~ms for the larger window, we can see in this $15$~min granularity a few spikes with reported latencies larger than $70$~ms ($2\times$ the median) each persisting for roughly $15$ seconds or more. The first large spike starting at the $523^{rd}$ second of the window lasts for $30$ seconds. We echo the popular speculation~\cite{15sec_1, 15sec_2} that such high latencies occur due to sub-optimal hand-offs that last for multiples of $15$ seconds. Interestingly, there are many smaller spikes throughout with peak latencies around $50$~ms, and the ping latencies change continually reflecting the LEO dynamics. Such highly dynamic behavior might translate to transport and application performance bottlenecks. We also observe transient packet loss happening often when the spikes start/end.

The latency time series analysis sheds light on the unique implications of LEO dynamics: frequent hand-offs could lead to a user terminal being mapped to a satellite such that the latencies are inflated persistently for the entire duration that the terminal is connected to that satellite. Given satellite trajectories are publicly available~\cite{celestrak}, can we predict LEO's dynamicity? To put it in a different way, can we build LEO performance predictors that could provide useful insights to applications to tackle such predictable dynamicity? We explore this in \S\ref{subsec:leo_prediction}.


\subsection{Dissecting LEO's bent-pipe latency} 


Before building a LEO performance predictor based on the known orbital dynamics of Starlink, it is imperative to ascertain that the performance is predominantly influenced by the LEO component which is the satellite bent-pipe connection, rather than the terrestrial wide area Internet element, within the end-to-end path. The end-to-end path between Starlink terminals and the Internet services can be broken down into distinct segments which we designate as S$1$ through S$6$ as in Fig.~\ref{fig:leo_path}. S$1$ is the host-to-Starlink terminal component, S$2$ is the satellite bent-pipe, S$3$ -- S$5$ are Starlink's internal terrestrial components (verified by mapping IP addresses to ASNs), and S$6$ is the wide area Internet part of the path.

To gain insights into how each of these segments contributes to the end-to-end latency, we run traceroute experiments from all measurement clients to Google’s anycast DNS server ($8$.$8$.$8$.$8$). These traceroutes run at a cadence of $1$~s for $2$ hours, $3$ times a day, over a span of $7$ days. We plot the segment-wise one-way median latency in Fig.~\ref{fig:latency_bar_plot}. The LEO bent-pipe component of the path (S$2$) contributes the largest component to the end-to-end path latency -- $54\%$ for CMG, $65\%$ for EDH, $63\%$ for LDN, $52\%$ for BCN-$1$, and $59\%$ for BCN-$2$ at the median.

\subsection{Some terminal orientations could be worse!} 
\label{subsec:terminallatency}
\begin{figure}[t]
  \centering
  \subfigure[]{\label{fig:grpc_azimuth_elevation}
    \includegraphics[width=0.4\textwidth]{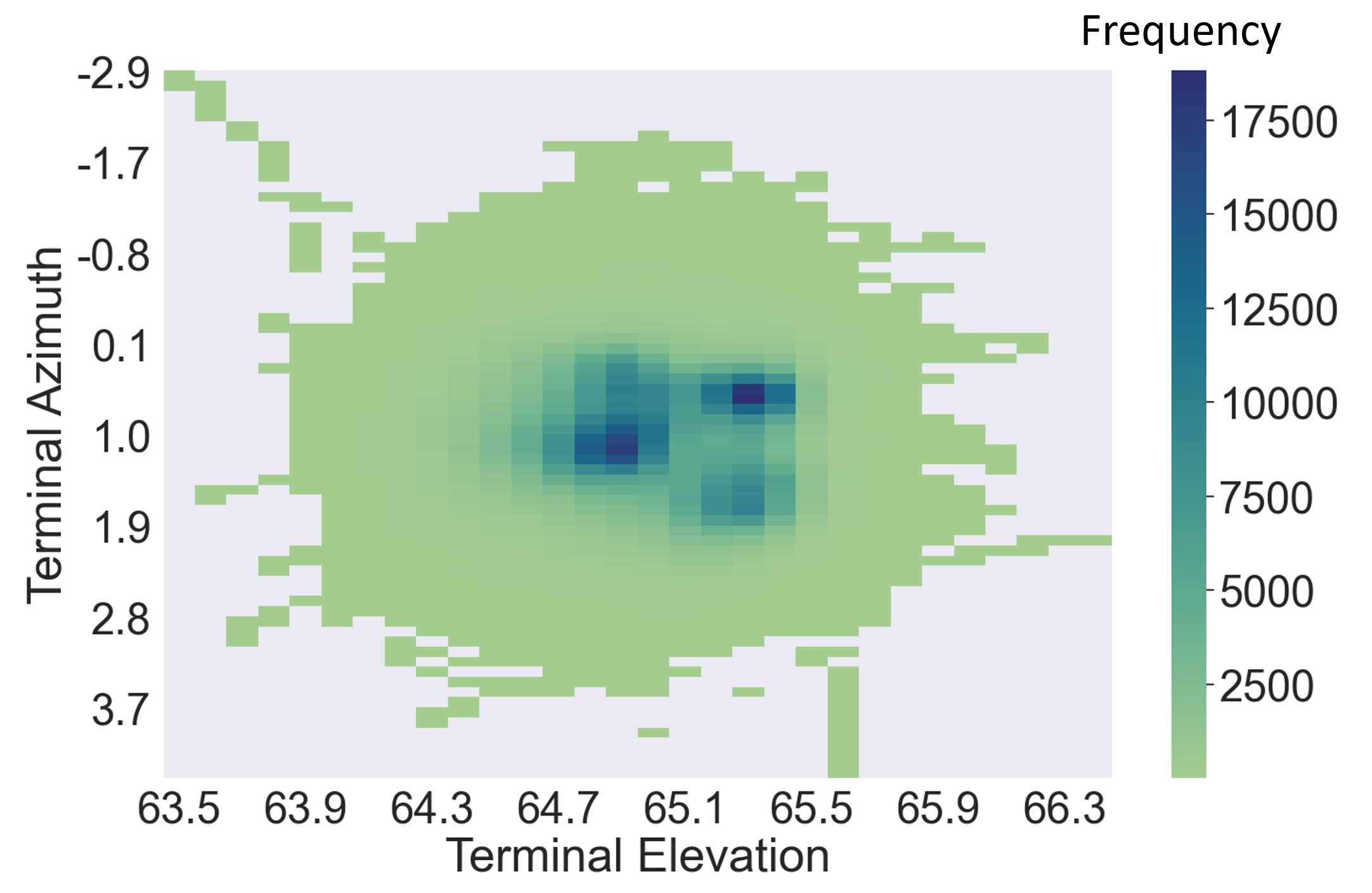}}
  \subfigure[]{\label{fig:grpc_orientation_latency}
    \includegraphics[width=0.4\textwidth]{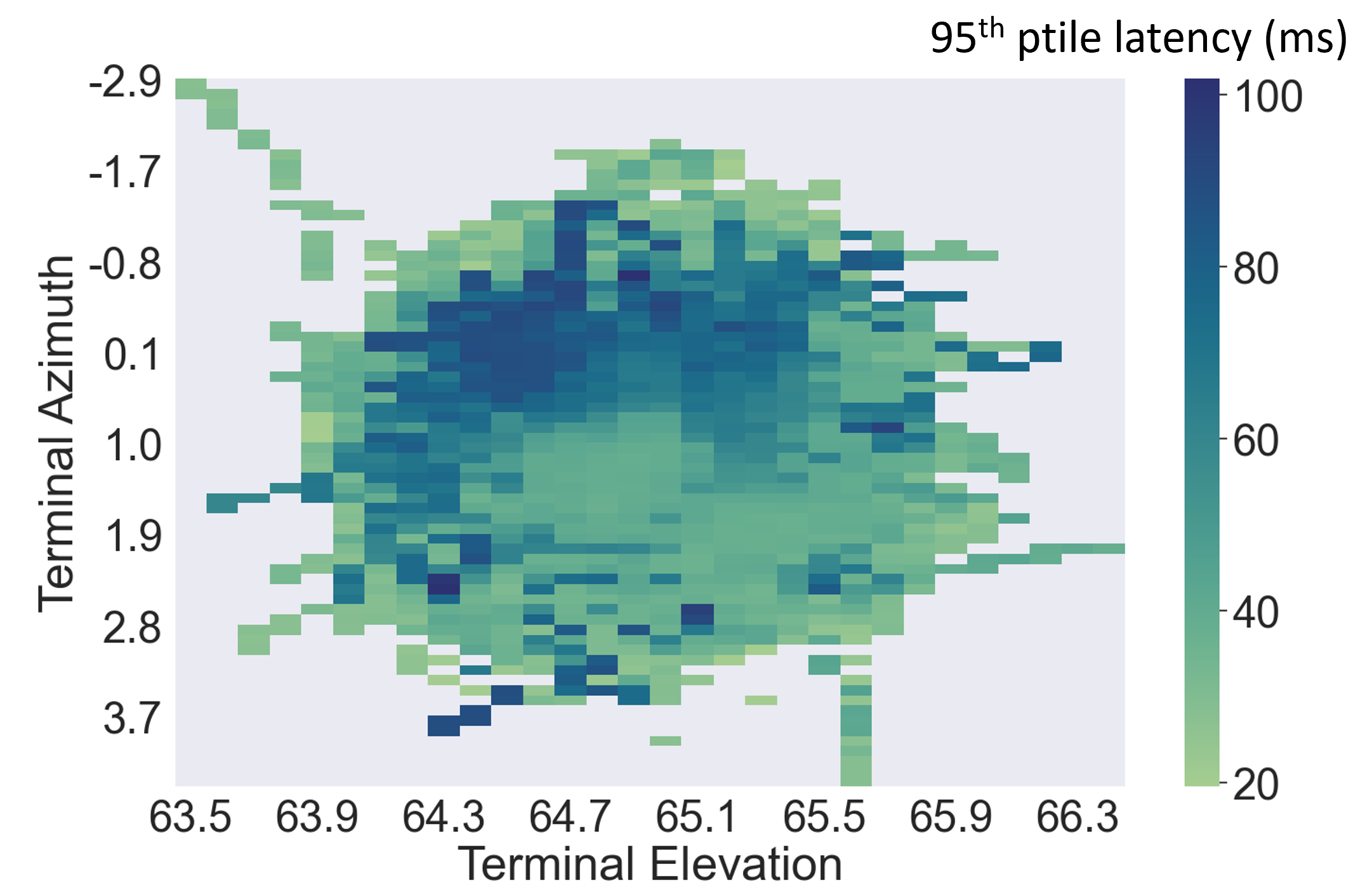}}
  \vspace{-0.15in}
  \caption{Heatmaps of (a) frequency of occurrence and (b) $95^{th}$ percentile latencies for combinations of antenna physical azimuth and elevation for the measurement client in Barcelona for $320+$ hours at a cadence of $1$ second.}
  \label{fig:orientation}
  \vspace{-0.2in}
\end{figure} 
The Starlink terminals at the measurement clients have an inbuilt telemetry system that continually monitors the physical orientation of the terminal (azimuth and elevation) along with packet drop rate to PoP (point of presence on the other side of the bent-pipe last mile), ping latency to PoP, etc. To understand how/if user terminal orientation impacts observed latency, we collect terminal telemetry data for $320+$ hours from the BCN-$1$ node at a granularity of $\sim${}$1$ second.

The physical orientation of the user terminal (antenna) seldom changes -- $90\%$ of the azimuth and elevation values lie within narrow bands of [$0.2$, $1.8$] (possibly to avoid GEO-arcs) and [$64.5$, $65.4$] degrees  respectively. 
Fig.~\ref{fig:grpc_azimuth_elevation} shows a few clusters of popular azimuth and elevation combinations with a long tail of other values. Note that the antenna changes its azimuth and elevation gradually, albeit at a slow median pace of $0.0001\degree$ per second for both axes. We do not have visibility into the phased-array beam~\cite{phased_array_unknown} orientation on top of the physical orientation of the antenna, but as we see next, the physical orientation could have important network performance implications.

Fig.~\ref{fig:grpc_orientation_latency} plots the heatmap of $95^{th}$ percentile latencies for different combinations of terminal azimuth and elevation. The deepest hues (largest values of latency; still less than $100$~ms though) correspond to less popular orientations (see Fig.~\ref{fig:grpc_azimuth_elevation}). \db{We observe similar trends for median (instead of $95^{th}$ percentile) latencies too.} This interesting observation, alongside the persistent latency spikes that we see in Fig.~\ref{fig:latency_time_series}, and the observation in Fig.~\ref{fig:latency_bar_plot} that the bent-pipe adds the largest component to the end-to-end latency, entice us to believe LEO network performance could be predicted to a reasonable degree. Satellite trajectories are predictable over time: publicly available satellite TLEs (two-line elements)~\cite{celestrak} define these trajectories for the next tens of hours and are updated regularly. Hence, the delay component driven by the satellite trajectories and gateway locations could be predictable.

\db{Note that the most frequent antenna orientations in Fig.~\ref{fig:grpc_azimuth_elevation} do not offer the lowest latencies in Fig.~\ref{fig:grpc_orientation_latency} though. We do not claim that the antenna orientations are geared toward lower latency, rather we claim some of the infrequent orientations could lead to latency inflation.}

\greybox{\textbf{\S\ref{sec:latency} Takeaway:} Starlink-like LEO networks could offer low ($30$-$50$~ms) latency at the median but are subject to very high tail latencies. Such high latencies are a result of a complex mix of constellation geometry, orientation and location of the user terminal, deployment of gateways, satellite hand-offs, as well as other unforeseen variable factors in the terrestrial path segments.}

%% file: video-streaming-predictor.tex
\section{\underline{P}redictors: Toward infusing applications with LEO awareness}
\label{sec:predictors}



\label{subsec:leo_prediction}

As we observe in the previous section, while Starlink LEO offers reasonable median latencies of few tens of milliseconds, the tail values could be an order of magnitude higher. Such high latencies are due to persistent inflated latency periods (probably due to sub-optimal hand-offs), sub-optimal antenna positioning, etc. This, accompanied by high ($\sim${}$2\%$-$5\%$) packet loss~\cite{kassem2022browser, starlink_loss} in LEO, makes it difficult for upper layers in the stack to adapt to changing network conditions over LEO. For example, both loss and delay-based congestion control schemes suffer as these signals become noisier due to non-congestive latency inflation and loss. Applications like video streaming, which rely on throughput predictions, could suffer due to such high dynamicity that happens naturally in these networks. Given, satellite position information is publicly available~\cite{celestrak} and the Starlink terminals expose a rich interface to access terminal orientation data, continuous telemetry, etc., we explore the opportunity to \textit{predict} LEO performance (latency and throughput) in this section. The predictors, as part of the T$3$P stack, could offer useful signals to the upper layer to adapt on. We  demonstrate how our custom-built predictor improves video streaming quality of experience.

\subsection{Internals of the LEO predictors}

\parab{Algorithms:} For the latency and throughput predictor models, we tried $3$ algorithms -- XGBoost~\cite{xgboost} (Extreme Gradient Boosting), 
LSTM~\cite{sherstinsky2020fundamentals} (Long Short-Term Memory), and ARIMA~\cite{arima} (AutoRegressive Integrated Moving Average). XGBoost is a gradient-boosted decision tree-based algorithm widely used for classification and regression. We use XGBoost for this regression problem as it effectively handles non-linear relationships between the features (we describe below) and the target variable (latency/throughput). Moreover, it is an ensemble learning method that combines the predictions of multiple decision trees, leading to improved accuracy and stability. LSTM is a specialized recurrent neural network that is adept at learning and remembering dependencies in sequential data, making it highly suitable for historical data (like latency) and its correlation with the features. ARIMA, on the other hand, is a simple time-series analysis technique that is particularly useful for handling stationary data and data with seasonality. In our experiments, ARIMA offered significantly lower accuracies than both XGBoost and LSTM. Hence, we refrain from discussing ARIMA further.


\parab{Features:} We use the following features to train the predictors -- $1$. the terminal location coordinates (latitude, longitude, altitude), $2$. publicly available satellite trajectory information that gives the instantaneous locations (azimuth, elevation, and distance from the user terminal) of all satellites within the field of view of the user terminal, $3$. the user terminal orientation data (azimuth \& elevation),  $4$. the historic (last $5$ seconds) performance (latency/throughput) data recorded by the user terminal, and $5$. the time (timestamp in second). 

\parab{Experimental setup:} Across $2$ measurement client locations (BCN-1 in Spain and EDH in the UK), we train the models on $\sim${}$19$ hours of data each at a granularity of $1$ second and test them with $\sim${}$5$ hours of data at the same granularity. We predict end-to-end latencies to servers hosted on geographically closest Azure locations.

\parab{Results - Latency:} Across both locations, LSTM demonstrates superior accuracy, as seen in Table.~\ref{table:predictor_factor}, compared to the XGBoost model. It achieves a mean accuracy of $\sim${}$96\%$ and performs consistently better than XGBoost at the median, $95^{th}$, and $99^{th}$ percentiles. When comparing the Relative Root Mean Square Error (RMSE) between XGBoost and LSTM models, LSTM’s superiority is evident. Its RMSE remains below $2.27$~ms across all locations, whereas the XGBoost model reaches a maximum RMSE value of $5.6$~ms for a link with a mean latency of $32.46$~ms.

\parab{Results - Throughput:} We trained similar models for predicting throughput, though the accuracies of these models are relatively lower (throughput is \textit{hard} to predict \cite{predict_survey}, especially given the LEO dynamics). We were able to achieve maximum accuracy of $73.92\%$ for LSTM and $80.80\%$ for the XGBoost model. In contrast to the latency predictors, in the case of the throughput predictors, we observe XGBoost to be performing better than the LSTM model, as seen in Table.~\ref{table:predictor_factor}. This is because XGBoost is based on decision trees and is effective in capturing complex interactions between features, which is useful for predicting throughput that usually depends more on a complex calculus involving various factors. Latency movements on the other hand are more continuous, in general, and are predicted well by LSTM which is particularly effective in capturing the temporal dependencies in sequential data.



\begin{table*}[t]
\centering
\resizebox{\textwidth}{!}{%
\begin{tabular}{c|cccc|cccc|cccc|cccc}
\hline
     & \multicolumn{8}{c|}{Throughput Predictors}  & \multicolumn{8}{c}{Latency Predictors} \\ \hline
\textbf{ Testbed}    & \multicolumn{4}{c|}{XGBoost}  & \multicolumn{4}{c|}{LSTM} & \multicolumn{4}{c|}{XGBoost} & \multicolumn{4}{c}{LSTM}\\  
\textbf{ Location}      & MAPE    & RMSE & $5\%$ & $10\%$  & MAPE    & RMSE & $5\%$ & $10\%$ & MAPE     & RMSE & $5\%$ & $10\%$  & MAPE    & RMSE & $5\%$ & $10\%$ \\  \hline
\textbf{BCN-1} & 19.2\% & 190 kbps & 84.99\% & 92.92\% & 26.08\% & 153 kbps & 40.47\% & 61.33\% & 10.68\% & 5.53 ms &36.52\% & 61.91\% &  3.85\% & 2.03 ms & 75.10\% & 93.55\%\\
\textbf{EDH} &  41.67\% & 291 kbps & 9.02\% \% & 17.20\% & 87.34\% & 456 kbps & 1.13\% & 2.32\% & 9.95\% & 5.60 ms & 37.55 \% & 65.00\% & 3.65\% & 2.27 ms & 76.38\% & 94.51\% \\
\hline
\end{tabular}}
\caption{Comparison of the accuracy of different prediction models across two locations in Spain and the UK. MAPE stands for Mean Average Percentage Error, RMSE represents the Root Mean Squared Error, and the $5\%$ and $10\%$ column headers represent the percentage of predicted values that lie within $ \pm5\%$ and $ \pm10\%$ range of the actual values, respectively.}
\label{table:predictor_factor}
\end{table*}

\subsection{Augmenting applications with LEO predictors}

Can we use these predictors to improve the performance of applications running over LEO networks? Toward understanding this broad opportunity, we run a simple video streaming experiment in $2$ of our deployment sites -- BCN-$1$ in Spain, and EDH in the UK. We generate network traces using \texttt{iperf} and the terminal-generated data for $10$ hours at a granularity of $1$ second and use a similar setup as in \cite{pensieve_qoe} to run our video streaming tests -- an emulated LEO network over Mahimahi \cite{netravali2015mahimahi}, and a custom encoded video from Pensieve of $3$~min duration with $6$ qualities. The encoded bitrate ratio is $\sim${}$1.5$ between adjacent qualities and tracks. We use RobustMPC~\cite{yin2015control} (control-theoretic) as the ABR (adaptive bitrate) algorithm and measure the general video streaming QoE metric~\cite{pensieve_qoe} (combines high bitrate, minimal rebuffering, and smoothness) across $100$ runs of each algorithm as described below.


We use $3$ variants of the MPC algorithm -- $1$. Default MPC (MPC-D): the default implementation that uses its own custom throughput predictor, $2$. LEO-aware MPC (MPC-L): uses the LEO throughput predictor (XGBoost) we trained with client-local data, and $3$. MPC with Oracle (MPC-O): knows the throughput trace in advance. Across both locations and multiple runs, we observe MPC-O (a hypothetical upper-bound) outperforms the other schemes (improves QoE by $30\%$ at the median over the baseline MPC-D) consistently, which is along the lines of expectation. Interestingly though, MPC-L, backed by our custom-built LEO throughput predictor, also consistently outperforms MPC-D by $25\%$ at the median and $23\%$ at the $95^{th}$ percentile.


\greybox{\textbf{\S\ref{sec:predictors} Takeaway:} LEO predictors could offer useful signals to upper layers. Applications, like video streaming, that rely on predicting network performance could significantly improve QoE by adopting the T$3$P stack. Since these adaptations are application-specific, we are able to optimize for application-specific requirements, such as QoE for video streaming.} 

%% file: bbr2_parameter_tune.tex
\section{\underline{T}ransport: Fine-tuning congestion control over LEO}
\label{sec:transport}

The Internet transport is evolving from an era of loss-based congestion control (CC) schemes~\cite{ha2008cubic, floyd2004newreno} toward delay-based ones~\cite{bbr, cardwell2019bbrv2, copa, dong2018pcc}, most of which rely on micro-experiments to identify the right sending rates. How do such CC schemes fare over the SpaceX Starlink LEO network, given network latencies change continually due to satellite motion and occasional hand-offs? In this section, we quantify the performance (and shortcomings) of CC schemes over Starlink LEO, explore the joint-optimization of $2$ intuitive parameters of BBRv$2$ which clearly falls short of expectations otherwise, and briefly comment on the fairness of such optimized variants to Cubic flows.

\db{Recent work~\cite{michel2022first} on measuring Starlink performance assumes a fixed CC scheme (Cubic) for all the experiments. Given the low latency promise~\cite{mark_handley_hotnets2018} of these networks, Cubic might not be the most efficient choice -- it is known to fill up buffers thus inflating end-to-end latencies. Another recent study~\cite{kassem2022browser} briefly compares different CC performances over LEO, finding that BBRv$1$ offers higher throughput than others. We rather focus on BBRv$2$, a more recent BBR version, which struggles to perform over the Starlink network.}

While the previous section leveraged raw signals of predictable LEO dynamicity to generate more sophisticated network performance signals that could improve application performance, in this section, we take a somewhat complementary stance -- augmenting existing end-to-end transport with LEO intuition. Note that these are different choices for application providers available in the T$3$P stack (check \S\ref{subsec:t3p_stack}) toward building LEO-aware applications.


\begin{figure*}[t]
  \centering
  \subfigure[]{\label{fig:cc_comparison}
  \includegraphics[width=0.35\textwidth]{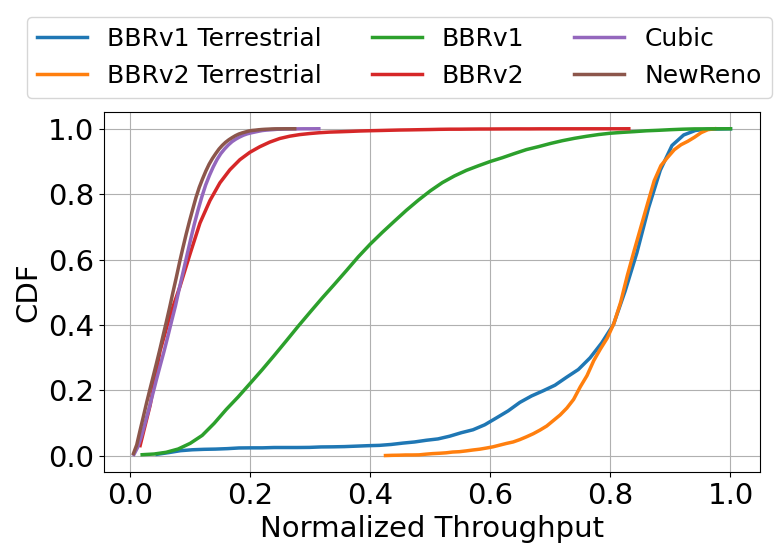}}
  \subfigure[]{\label{fig:bbrv2_1}
    \includegraphics[width=0.35\textwidth]{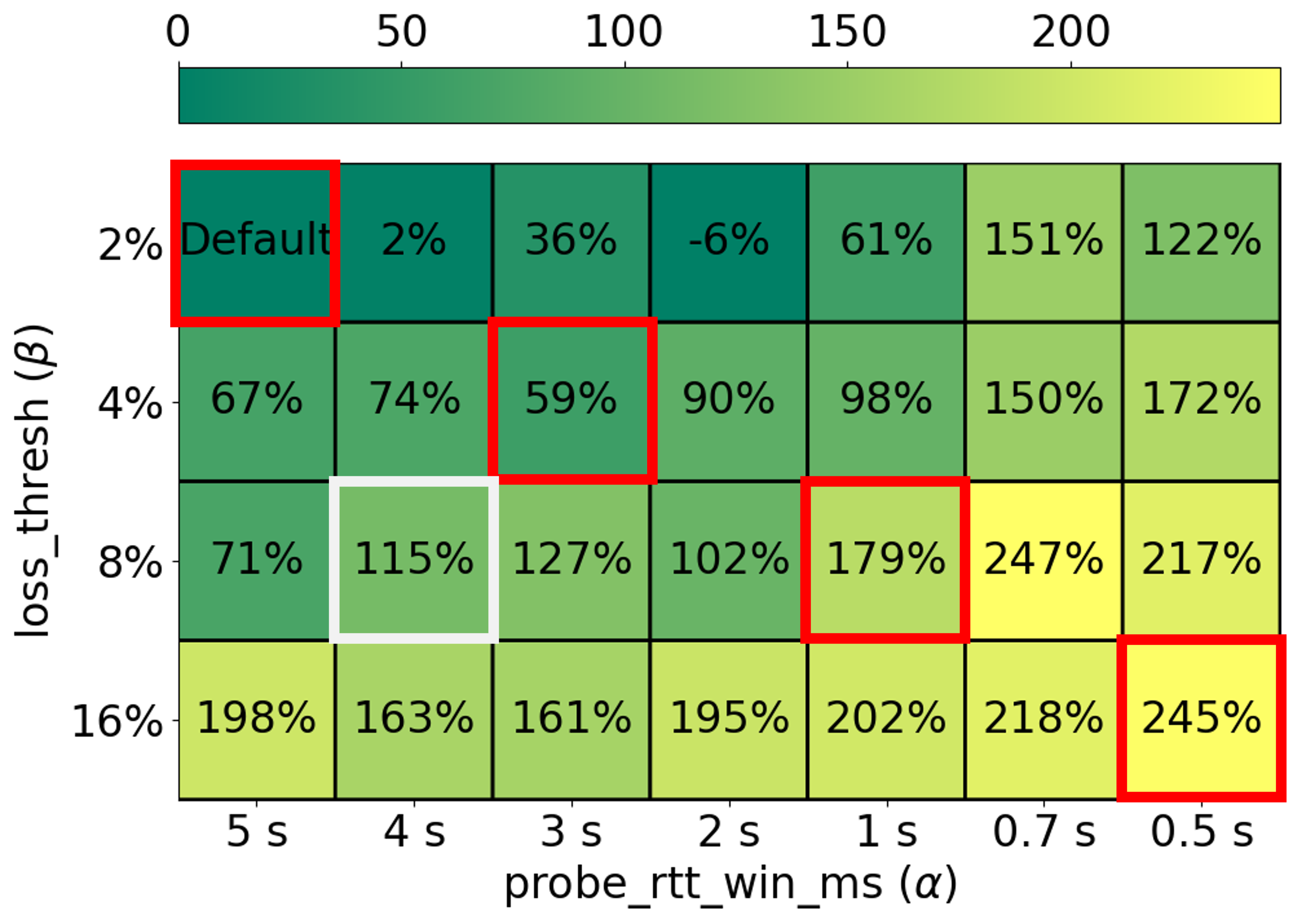}}\\
  \subfigure[]{\label{fig:bbrv2_2}
    \includegraphics[width=0.35\textwidth]{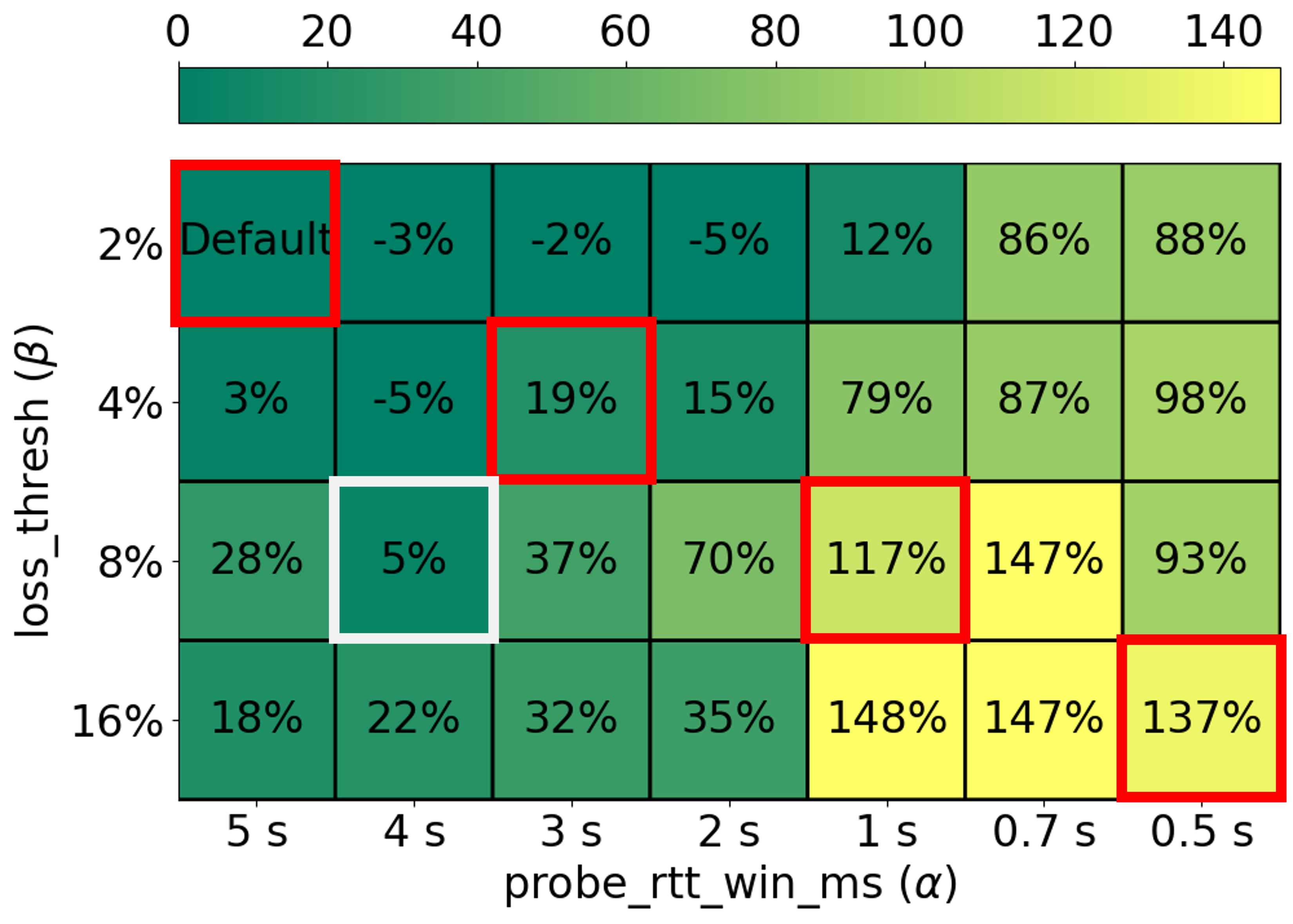}}
  \subfigure[]{\label{fig:fair}
    \includegraphics[width=0.35\textwidth]{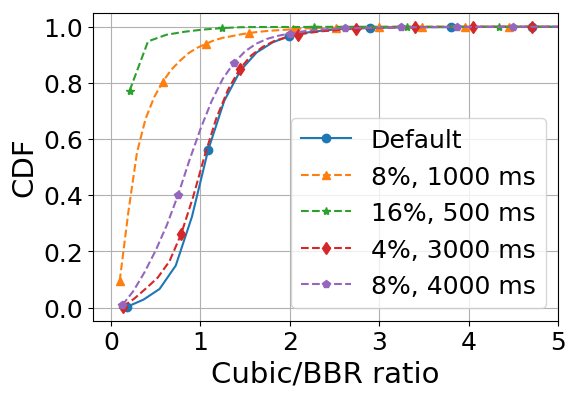}}  
  \caption{CC measurements from BCN-$1$ to an Azure server in Netherlands. (a) CDFs of throughput at $1$~s granularity for various CC schemes; normalized by the maximum BBRv$1$ throughput (for terrestrial and LEO). (b) and (c) show the mean throughput percentage improvement and the $95^{th}$ percentile latency percentage inflation (resp.) heatmaps for different values of BBRv$2$ parameters. (d) CDFs of the ratio of throughputs of aggregated Cubic flows and aggregated BBRv$2$ flows for some of the parameter configurations (marked in red and white in (b) and (c)).}
\end{figure*} 

\subsection{Battle of the CC schemes}
At the BCN-$1$ node, we ran \texttt{iperf}~\cite{iperf} upload tests to an Azure-hosted measurement server in Netherlands\footnote{This cloud server is physically close to the BCN-$1$ node to make sure that the LEO bent-pipe (terminal-satellite-ground station) constitutes a significant component of the end-to-end path.} to compare the performance (throughput) of loss-based (Cubic~\cite{ha2008cubic}, NewReno~\cite{floyd2004newreno}) and delay-based (BBRv$1$~\cite{bbr}, BBRv$2$~\cite{cardwell2019bbrv2}) transport over Starlink and the terrestrial Internet. Note that BBRv$1$ is agnostic to packet loss while the more recent BBRv$2$ considers loss and ECN markings~\cite{ramakrishnan2001rfc3168} as signals of congestion alongside delay. We ran each CC scheme for $4$~min each time, $3$ times a day, for a total of $100$~min (per CC scheme) over both LEO and the terrestrial Internet. 
\db{All plots are at $1$~s granularity, thus translating to $6$,$000$ data points for each. Note that each $1$~s data point, in turn, translates to $\sim${}$30$ round-trips in our setup.}


In Fig.~\ref{fig:cc_comparison}, we plot the CDFs of throughput for CC schemes 
normalized with the maximum observed throughput for the terrestrial and LEO settings separately for the BCN-$1$ node. For the terrestrial setting, we only report BBR, as the loss-based schemes are known to operate close to the capacity. We observe that CC performance over LEO is significantly worse than over the terrestrial Internet. This is in line with observations in prior simulation-based work~\cite{kassing2020exploring} and a browser-plugin-based study~\cite{kassem2022browser}. While terrestrially both BBRv$1$ and BBRv$2$ could achieve median throughput of $82\%$ the maximum capacity (maximum observed throughput), over Starlink LEO they could only achieve $32\%$ and $8\%$, respectively. Also, the loss-based schemes, Cubic and NewReno, perform significantly worse, with median throughput reaching only $8\%$ and $7\%$ of the capacity respectively. Note that this is contrary to results from previous simulations~\cite{kassing2020exploring}, and is likely due to the high loss rates~\cite{kassem2022browser} alongside latency variations over Starlink LEO. We perform this experiment for all measurement clients (and similar setups with nearby servers) and observe similar trends. The interesting and somewhat counter-intuitive observation that BBRv$2$ performs significantly worse than the older BBRv$1$ (due to its greater sensitivity to loss) pushes us to jointly tune $2$ relevant configuration parameters of BBRv$2$ toward improving performance over these highly dynamic networks.


\subsection{Pushing BBRv$2$ beyond the default} 
BBR and other rate-based CC schemes regularly probe the network (bandwidth, minimum round trip time or RTT) and adapt the sending rate and congestion window accordingly. BBRv$2$ has $2$ distinct modes in its steady state -- $1$. \textit{ProbeBW}: probes the link bandwidth at regular intervals by sending at a higher rate, and $2$. \textit{ProbeRTT}: probes the RTT, if it is not updated recently, by sending at a lower rate thus draining the queue. Both these distinct modes are governed by multiple configurable parameters that decide the frequency and intensity of the probes. We intuitively pick $2$ of these parameters, given the LEO setting, and jointly tune both by performing a coarse-grained $2$-dimensional grid search for this demonstration. These parameters are \textit{probe\_rtt\_win\_ms ($\alpha$)}, and \textit{loss\_thresh ($\beta$)}. $\alpha$ denotes the time interval between consecutive ProbeRTT mode entries. As Starlink latencies change continually, this is an intuitive knob to tune. $\beta$ denotes the loss rate threshold to stop probing for more bandwidth. Given Starlink is lossy~\cite{starlink_loss, kassem2022browser}, intuitively this is another interesting parameter to tune.


We perform a coarse-grained $2$ dimensional grid search on the ($\alpha$, $\beta$) parameter space, trying out $\alpha$ values that lead to more frequent RTT probes and $\beta$ values more robust to loss. We run the different combinations of $\alpha$ and $\beta$ for the BCN-$1$ uplink for $4$~min each, $3$ times a day, for a total of $100$~min. \db{At $1$~s granularity, this translates to $6$,$000$ data points per CC scheme. Each data point constitutes of $\sim${}$30$ round trips.} Across all measurements \db{($1$~s data points)} for a CC scheme, we compute the mean throughput and the $95^{th}$ percentile latency, and compare them with that of the default setting (reported in Fig.~\ref{fig:bbrv2_1} and \ref{fig:bbrv2_2} heatmaps). We observe that for lower values of $\alpha$ and higher values of $\beta$ we get better throughput than the default, along the lines of our intuition, but often at the cost of increased latency (or RTT). For both the heatmaps, the bottom-right has warmer shades (higher values), signifying improved throughput at the cost of latency inflation. A close manual examination reveals that this is due to the costs and side-effects of probing. For example, probing RTT more frequently might also mean BBRv$2$ goes into the exponential startup phase more often when exiting the ProbeRTT mode. As a result, BBRv$2$ fills up the buffer, leading to increased queuing delay and thus an overall increased latency. Nevertheless, we find out interesting combinations of $\alpha$ and $\beta$ for which throughput is increased significantly with minimal inflation in latency. We define the \textit{best} performing configuration as the maximum throughput improvement with minimal increase in the RTT (i.e., $<10\%$). For BCN-$1$, the best configuration ($\alpha$=$4000$ ms and $\beta$=$8\%$) offers $115\%$ increase in the throughput over the default with only $5\%$ increase in RTT.


\parab{Toward BBRv$2$ spatial tuning} While the general trends across all locations are similar (warmer shades toward the bottom-right of the heatmaps), the \textit{best} parameter combinations (defined as the parameters which obtain the highest throughput improvements with $<10\%$ latency inflation over the default parameter values) for $\alpha$, $\beta$  vary across locations, with $\alpha = 5$,$000/4$,$000/2$,$000$~ms and $\beta = 8/16\%$ offering better operating points than the default. This observation opens up the possibility to tune BBRv$2$ differently across different locations based on the connectivity profile.


\parab{Fairness to Cubic} 
While a thorough analysis of the fairness of such parameter-tuned BBRv$2$ flows is left to future work, one significant question is whether the parameter modifications we advocate might affect fairness to legacy CC algorithms such as Cubic. For this purpose, we pick the default as well as three other $\alpha$, $\beta$ combinations (marked in red in Fig.~\ref{fig:bbrv2_1} and \ref{fig:bbrv2_2}) along the diagonal of the heatmaps, and the \textit{best} choice for BCN-$1$ (marked in white). For each combination, we run $8$ Cubic and $8$ tuned BBRv$2$ flows in parallel for $100$~min. Fig.~\ref{fig:fair} plots the CDFs of the ratio of aggregate Cubic throughput and the aggregate BBRv$2$ throughput at $1$~s granularity. We observe that as we move toward the bottom-right along the diagonal of the heatmaps, the ratios are lower in Fig.~\ref{fig:fair}, signifying BBRv$2$ becoming more aggressive. Nevertheless, the ratios with the \textit{best} choice $\alpha$, $\beta$ combinations follow the default CDF plot closely with only a nominal reduction in values. This demonstrates the opportunity to find good $\alpha$, $\beta$ combinations that do not significantly penalize legacy Cubic flows.


\greybox{\textbf{\S\ref{sec:transport} Takeaway:} Starlink LEO creates unique performance bottlenecks for both loss-based and delay-based CC schemes. BBRv$2$ could be potentially tuned to offer significantly higher throughput at the cost of minimal latency inflation.} 



%% file: testbed-2.tex
\section{\projectname{} Testbed}
\label{sec:testbeds}

We built a testbed, called \projectname{}, to run all the experiments over Starlink LEO network. The testbed:
\begin{itemize}
    \item Allows for orchestration and scheduling of experiments across measurement clients and servers.
    \item Gathers telemetry data that could be used to build a continuous LEO performance dashboard.
    \item Generates LEO link profiles \cite{leoscope-simulation-profiles}
    that could be used as inputs by various LEO simulators (like Hypatia~\cite{kassing2020exploring}) and emulators (like StarryNet~\cite{lai2023starrynet}).
    \item Allows Starlink users to volunteer/donate resources to the testbed without being affected while running personal applications. The LDN client in our experiments is one such node.
\end{itemize}

We also enabled useful triggers on \projectname{} based on various networking (latency, loss, etc.) and non-networking (weather conditions, satellite positions, etc.) conditions that could initiate experiments. Given the high dynamicity of Starlink LEO, this setup is useful to zoom in on interesting LEO-specific scenarios. 
\db{As part of this work, we have released the \projectname{} code~\cite{leoscope_code}.} The rest of this section touches upon \projectname{} design.


\begin{figure*}[t]
  \centering
  \subfigure[]{\label{fig:leopard_design}
    \includegraphics[width=0.75\textwidth]{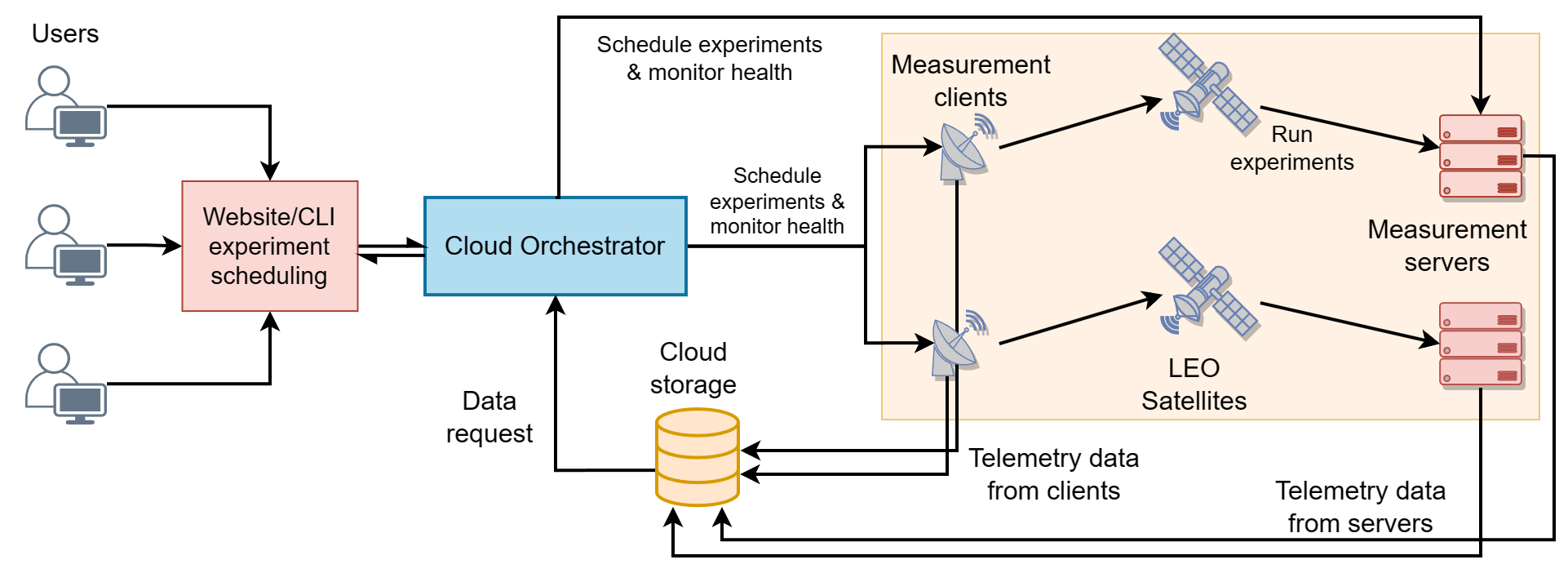}}
  \subfigure[]{\label{fig:leopard_terminal}
    \includegraphics[width=0.12\textwidth]{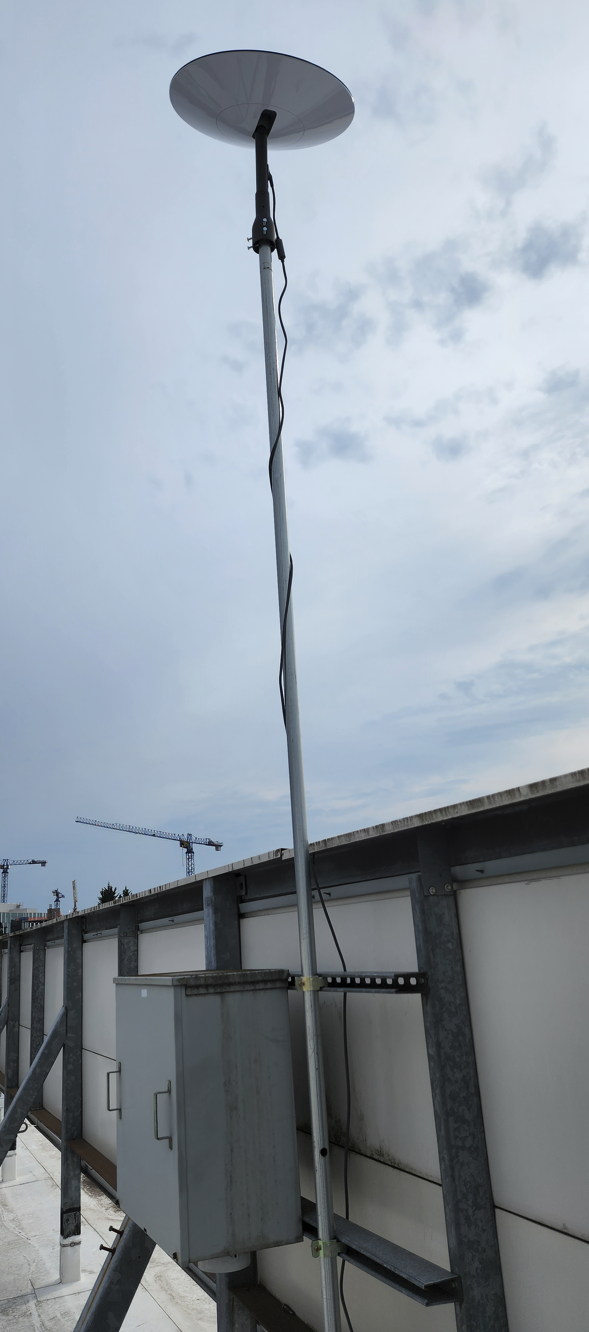}}
  \caption{(a) \projectname{} building blocks. (b) A user terminal deployment with a clear view of the sky.} 
\end{figure*}

\subsection{Broad design}
\label{sec:design}
Fig.~\ref{fig:leopard_design} explains \projectname{} in a nutshell. We touch upon the individual key components below.

\parab{Users} can submit network/application-level experiments to \projectname{}, request a schedule and a set of measurement clients and servers to run these experiments on, and view the results dashboard. 

\parab{Cloud orchestrator} is responsible for scheduling experiments across measurement clients and servers, managing schedule conflicts, and monitoring measurement node health. 

\parab{Measurement servers} are deployed on public cloud. For client-server measurement experiments like \texttt{iperf}~\cite{iperf}, measurement clients could reach out to these servers. The server daemon receives an experiment schedule from the cloud orchestrator, downloads the necessary artifacts, and sets up the local environment (if needed) before the client requests arrive. 

\parab{Measurement clients} consist of Starlink LEO user terminals with a reasonably clear view of the sky (see Fig.~\ref{fig:leopard_terminal}), and hosts (could be standard workstations, Raspberry Pi~\cite{rasp_pi}, or more capable edge devices like Azure Stack Edge~\cite{ase} or Intel NUC~\cite{intel_nuc}) behind the terminals that could access the Internet via the Starlink network. 

\parab{Cloud storage} receives measurement data from both measurement clients and servers. The storage is hierarchically structured into directories such that it is straightforward to navigate to a specific instance of an experiment on a measurement node. 

\parab{}Some engineering design choices we have made toward easy operations are:

\parab{Lightweight isolation} \projectname{} runs experiments on lightweight containers~\cite{container} that help automate experiment setup on measurement clients and servers and faster experiment startup. 

\parab{Kernel Service} \projectname{} exposes a firewalled kernel service for experiments that require low-level kernel-space configuration/modification (for example, testing kernel-space congestion control schemes and modifying interface-level queuing disciplines).

\subsection{\projectname{}'s unique features}
\label{subsec:unique}

Here we touch upon the unique \projectname{} features that help us dissect LEO network subtleties.

\parab{Ingesting user terminal data} The Starlink user terminals run continuous measurements in the background and collect bent-pipe latency, traffic volume, packet loss, antenna directional data, etc (as touched upon briefly in \S\ref{subsec:terminallatency}). These statistics are made available to the private network behind the user terminal via gRPC APIs~\cite{starlink_grpc}. \projectname{} clients continually gather this data to generate insights that enable the two following unique features of the testbed, discussed next.

\parab{Scavenger mode} This unique feature allows \projectname{} to locally (at the measurement client) preempt and reschedule experiments that result in upload/download traffic overhead when (higher-priority) user traffic is sensed leveraging the terminal-provided insights. This makes \projectname{} scalable -- volunteers with access to user terminals could join the testbed and contribute compute/connectivity free-cycles to the testbed. 
The LDN measurement client in the UK is run as a volunteer node on which we successfully tested scavenger mode.

\begin{figure}[t]
\begin{center} 
  \includegraphics[width=0.8\columnwidth]%
    {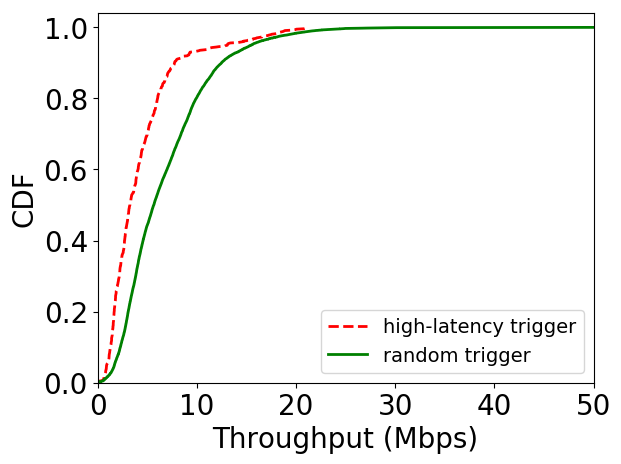}
  \caption{CDFs of BBRv$2$ throughput for BCN-$1$ uplink when triggers fire randomly versus on sensing inflated latency. The difference in median throughput is $43\%$, underscoring the utility of trigger-based scheduling in capturing LEO-specific events that impact the end-to-end performance of congestion-control schemes.}
  \label{fig:trigger-result-bbrv2}
\end{center}
\end{figure}

\parab{\underline{T}rigger-based scheduling} 
\projectname{} allows dynamic scheduling of experiments by specifying \textit{triggers} that execute experiments on the occurrence of diverse events of interest in the LEO context. For example, a trigger can be defined as follows: ``\textit{if current network latency $>$ $10\times$ the moving average (of network latency) of last ten windows, trigger traceroute probes to Google's $8$.$8$.$8$.$8$ service}''. Such trigger-based dynamic scheduling has $2$ key advantages: ($1$) Triggers allow the experimenter to pivot their measurements around specific network/physical events to capture interesting phenomena, more common in this dynamic LEO setting. Fig.~\ref{fig:trigger-result-bbrv2} shows how latency-sensitive triggers can be used to evaluate the performance of congestion-control schemes (in this case, BBRv$2$) when the LEO link latency is high. For the BCN-$1$ node, we run trigger-based BBRv$2$ (default config) upload experiments to an Azure measurement server in the Netherlands for a period of $7$~days, with the triggers firing randomly, versus when the latency jumps to a value at least twice the mean observed latency in the last $5$~s. The difference in median throughput obtained from the high-latency trigger schedule versus a random schedule is $43\%$ (with random runs offering a higher median and a longer tail), showing how triggers can help capture LEO-related events that impact congestion-control performance. Triggers can also be defined on publicly available information that can impact LEO link properties such as satellite trajectory information~\cite{celestrak}, local weather, etc. ($2$) Trigger-based schedules save data-transfer costs and storage space by confining the measurements to events of interest. For example, scheduling an experiment that runs a TCP flow for $24$ hours with an average throughput of $40$~Mbps incurs $\sim${}$3.5$~Tb of data-transfer costs and $\sim${}$138$Gb of storage space (only packet header captures) on a single measurement client. Defining a trigger-based schedule that restricts the run-time to $10\%$ of the total time can save $\sim${}$3$~Tb of data-transfer costs and $124$~Gb of storage. Triggers, therefore, contribute significantly to reducing operational costs. 

While we have used triggers to reduce the experiment footprint, they could be used to inform upper layers of interesting networking and physical events. The same functionality could be used to inform an application or transport about network latency elevation thus allowing them to take proactive actions toward optimizing the users' QoE. These triggers, due to their potential diverse usage, are offered as part of the T$3$P stack on \projectname{}.

%% file: discussions.tex
\section{Discussions and Future Work}
\label{sec:discussions}

This work sheds light on not only the LEO dynamics but also on the performance optimization opportunities that lie ahead. T$3$P or LEO awareness could be plugged into the network stack toward such optimizations that could impact the experience of the next billion users connecting to the Internet, cloud applications, and cloud services over the LEO broadband.

\subsection{Network stack optimisations}

\parab{Clean-slate transport} While we observe reasonable performance improvement by tuning BBRv$2$ over Starlink, we plan to explore clean-slate congestion control with LEO awareness baked into the design. Both loss and delay-based schemes run with the basic assumption that the minimum (propagation) delay in a network is fixed which does not hold true over LEO. An LEO-aware end-to-end transport should ideally be able to detect an LEO component on the path, accommodate dynamic minimum latencies, and consume predictions and/or triggers to augment its rate-finding algorithm.

\parab{Stochastic model for hand-off predictions} Although the latency and throughput predictors could offer useful LEO signals to application and transport, it could be useful to have stochastic models for hand-offs. Hand-offs could often be sub-optimal, resulting in persistent inflated latencies and/or transient non-congestive loss, thus necessitating applications to proactively react beforehand. For example, users of video conferencing applications report~\cite{reddit_stutters1, reddit_stutters2} occasional stutters over Starlink LEO. Application-level mitigation could be to consume the stochastic hand-off signals and reduce the video stream quality when the probability of a hand-off is high.

\parab{Latency and Throughput Prediction}
We are committed to improving the T$3$P stack by leveraging lightweight ML predictors or an ensemble of multiple predictors. We also plan to explore an online learning framework that quickly re-trains the models as  LEO network conditions change over time even at a given location (e.g., due to new satellite deployments, increase in number of users, etc.)


\parab{Multi-access connectivity} Several of our measurement clients have access to both LEO and terrestrial Internet. This could be more common in the future with Starlink adopters in the developed world using it as a secondary means of Internet connectivity. Note that the LEO and terrestrial fiber channels could offer drastically different offerings with the advent of inter-satellite lasers -- over longer distances, LEO offers lower latencies than terrestrial fiber. How could one design a transport that could take into account the diverse properties of the various access networks? Clearly, MPTCP-like~\cite{wischik2011design} solutions fall short as they assume access links to have similar offerings.

\parab{Understanding the physical layer} Starlink does not provide low-layer information, such as signal structure, Received Signal Strength (RSS), the satellite that the terminal is connected to, and the used channel. \cite{todd-inference} proposes a blind signal identification approach to infer the synchronization sequences in the Starlink downlink signal. One can potentially develop approaches to estimate RSS based on the inferred synchronization sequences (e.g., using correlation).

\begin{figure*}[t]
\begin{center} 
  \includegraphics[width=0.67\textwidth]%
    {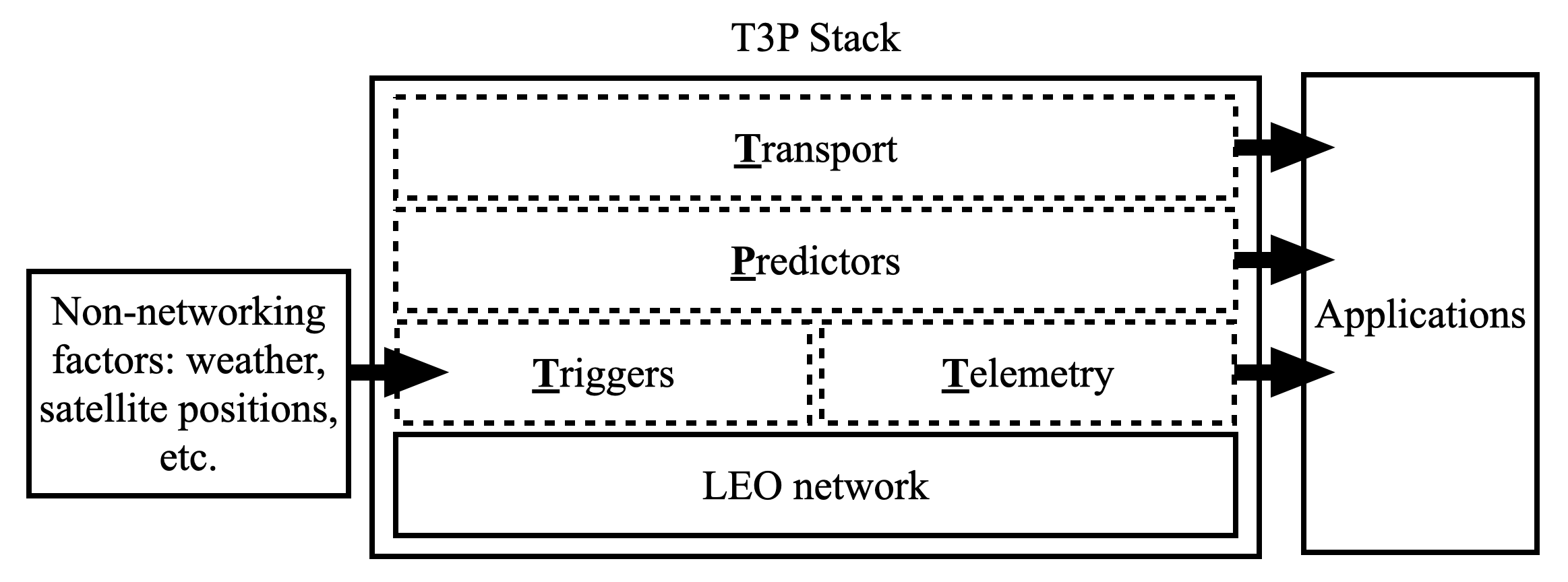}
  \caption{The LEO-aware T$3$P stack could offer rich signals to applications.}
  \label{fig:t3p_stack}
\end{center}
\end{figure*}

\subsection{Toward an LEO awareness library} 
\label{subsec:t3p_stack}
We anticipate that all the above performance optimization opportunities could build on one or more of T$3$P components. Hence, we envision the T$3$P stack, as shown in Fig.~\ref{fig:t3p_stack}, as an inherent building block of \projectname{} that could not only help run various network and application-level experiments on \projectname{} but also offer rich LEO-awareness signals to the upper layers (transport and application). 
Applications will have the flexibility to consume raw telemetry and/or triggers or more sophisticated predictions built on top of the raw signals.


\subsection{Future of \projectname{}}
As we have effectively used \projectname{} to run our experiments over Starlink LEO across multiple locations with ease, we plan to release \projectname{}. The T$3$P offerings are inherently baked into \projectname{} -- adopters could build experiments on top of these LEO-awareness signals or leverage the same to augment application performance. Also, \projectname{} runs continuous measurements across multiple locations, generating real LEO link profiles \cite{leoscope-simulation-profiles} in the process, which can act as important inputs for a large body of simulation~\cite{kassing2020exploring} and emulation work~\cite{lai2023starrynet} in this space.

\section{Related Work}


A large body of work from the nineties~\cite{lwood_thesis, bau2002topologies, chan2000optical, kwok2001cost, evans1998satellite, werner1995analysis, chan1999optical, wu1994mobile, kiesling1990land, comparetto1994global, sterling1991iridium, wiedeman1992globalstar, akturan1997path, maral1994ways} explores various networking aspects of the previous generation satellite networks. A \textit{lot} has changed since then -- the scale of deployments, the LEO broadband goals, coverage, performance expectations, and the way today's Internet protocols work. With this new wave of LEO satellite mega-constellations, the networking community has quickly realized that most of these past works need to be re-evaluated and re-calibrated in this new context. Three papers~\cite{bhattacherjee2018gearing, handley2018delay, klenze2018networking} in HotNets'$18$ set the path for such explorations in this `new space' context and have led to a flurry of interesting work in the last few years. While such work includes LEO network design~\cite{bhattacherjee2019network}, inter-domain routing~\cite{giacomo_ccr_ibis}, satellite downlink scheduling~\cite{vasisht2021l2d2}, etc., Hypatia~\cite{kassing2020exploring} was an early effort to deeply understand these networks by leveraging packet-level simulations. While it effectively captures the LEO dynamicity in its simulations, it lacks the realism of the actual LEO networks that are impacted by various factors like weather, ground station locations, link modulation, etc. This work was also the first to shed light on the CC performance bottlenecks over these dynamic LEO networks. Another work~\cite{kassem2022browser} leverages a browser plugin to collect insights on Starlink performance and passively analyze collected traces. In our work, we rather deploy a real Starlink testbed (unlike Hypatia's simulations) and run active end-to-end experiments. A more recent work, StarryNet~\cite{lai2023starrynet}, emulates integrated space-terrestrial networks. \projectname{} generates link profiles~\cite{leoscope-simulation-profiles} from real telemetry that could be treated as inputs to this and similar emulators.

Past works on satellite CC~\cite{caini2004tcp, westwood, peach} primarily focus on the GEO (Geostationary Orbit; $35$,$786$~km) satellites which experience high latencies due to the long bent-pipes. TCP Hybla~\cite{caini2004tcp}, although a good fit for such large RTTs, is not relevant in the context of these LEO networks that are characterized by reasonable RTTs but with large fluctuations. Our work, instead, explores CC schemes that dominate today's Internet in this new context.

We built \projectname{} by adopting learnings from existing testbeds like PlanetLab~\cite{planetlab2023web}, M-Lab~\cite{dovrolis2010mlab}, RIPE Atlas~\cite{atlas2023web}, BISmark~\cite{sundaresan2014bismark}, SamKnows~\cite{samknows2023web}, ORBIT~\cite{Orbit}, PhoneLab~\cite{PhoneLab}, and Monroe~\cite{alay2017monroe}. \projectname{} offers unique features useful for the LEO setting and integrates the T$3$P offerings that leverage LEO network signals.